\documentclass[aps,prd,showpacs,nofootinbib,preprintnumbers,amsmath,amssymb,mcite,superscriptaddress,twocolumn]{revtex4-1}

\usepackage[table]{xcolor}
\usepackage{graphicx}
\usepackage{dcolumn}
\usepackage{bm}
\usepackage{esvect}
\usepackage[latin1]{inputenc}
\usepackage[T1]{fontenc}
\usepackage{accents}

\usepackage{hyperref}
\hypersetup{colorlinks=true,linkcolor=magenta,anchorcolor=green,citecolor=cyan,filecolor=black,menucolor=black,urlcolor=brown}
\usepackage{slashed}
\newcommand{\be}{\begin{equation}}
\newcommand{\ee}{\end{equation}}
\newcommand{\ba}{\begin{eqnarray}}
\newcommand{\ea}{\end{eqnarray}}

\newcommand{\V}[1]{\vec{#1}}
\renewcommand{\d}{\partial}
\newcommand{\D}{\mathrm{d}}
\newcommand{\f}[2]{\frac{#1}{#2}}

\newcommand{\munu}{{\mu\nu}}

\newcommand{\h}{\f{1}{2}}

\newcommand{\Bmat}{\begin{matrix}}
\newcommand{\Emat}{\end{matrix}}
\newcommand{\Barr}{\begin{array}}
\newcommand{\Earr}{\end{array}}

\newcommand{\beq}{\begin{equation}}
\newcommand{\eeq}{\end{equation}}
\newcommand{\beqn}{\begin{eqnarray*}}
\newcommand{\eeqn}{\end{eqnarray*}}

\begin{document}

\title{Classical Casimir pressure in the presence of axion dark matter}

\author{Philippe Brax}
\affiliation{Institut de Physique Th\'eorique, Universit\'e  Paris-Saclay, CEA, CNRS, F-91191 Gif-sur-Yvette Cedex, France}
\author{Pierre Brun}
\affiliation{Irfu/D\'epartement de Physique des Particules, Universit\'e  Paris-Saclay, CEA, F-91191 Gif-sur-Yvette Cedex, France}

\begin{abstract}
We study the effects of an oscillating axion field  on the  pressure between two metallic plates. We consider the situation where a magnetic field parallel to the plates is present and show that the electric field induced by the coupling of the axion to photons leads to resonances. When the boundary plates are perfect conductors, the resonances are infinitely thin whilst they are  broadened when the conductivity of the boundary plates is taken into account. The resonances take place at  the tower of distances close to $d_n=\frac{(2n+1)\pi}{m}$
where $m$ is the axion mass and have a finite width and height depending on the conductivity. The resulting resonant pressure on the plate depends on the induced polarisation at the surface of the plates.  We investigate the reach of future Casimir experiments in terms of the axion mass  and the conductivity of the boundary plates. We find that for large enough conductivities, the axion-induced pressure could be larger than the quantum Casimir effect between the plates.
\end{abstract}

\maketitle


\section{Introduction}

The standard model contains a few naturalness issues, one of them being the so-called strong-CP problem. This is an archetypical fine tuning problem : QCD (Quantum Chromodynamics) is observed to respect the CP symmetry, as verified at the  $10^{-10}$ level by the non-observation of the neutron EDM (Electric Dipole Moment)~\cite{Pendlebury:2015lrz}. In the strict context of the standard model, QCD contains a CP-violating $\theta G\tilde G$ term, where $G_{\mu\nu}$ is the gluon field strength and no degree of freedom allows for the $\theta$ term to relax to zero. The Peccei-Quinn mechanism consists in introducing a new U(1) symmetry that is broken spontaneously at some high energy $M$, making the $\theta$ term dynamical and solving  the above puzzle~\cite{PQMechanism,PhysRevD.16.1791}. This is done at the expense of introducing new fields, {\it e.g.} new heavy fermions~\cite{KSVZ1,KSVZ2}, or extending the Higgs sector~\cite{DFSZ1,DFSZ2}. Eventually, there remains a single degree of freedom corresponding to  a massless pseudoscalar goldstone boson called the axion~\cite{WeinbergAxion,Wilczek:1977pj}. Because of the QCD phase transition, $\langle q\bar{q}\rangle\neq 0$ three things happen : the axion acquires a small mass, the $\theta$ term can effectively relax to zero, and the axion $\phi$ acquires an effective interaction with the electromagnetic field, of the form $\phi /M \ F\tilde{F}$, where $F_{\mu\nu}$ is the electromagnetic field strength.

Independently, observations of the universe at all scales from dwarf galaxies to the
whole Hubble radius favour the presence of a new type of cosmological fluid, which we call cold dark matter~\cite{Workman:2022ynf}. The energy density of dark matter today represents about one third of the total energy, and 84\% of the total mass~\cite{planck2018}. The axions could well be the source of dark matter. They could  have been produced in the early universe during the Peccei-Quinn symmetry breaking in a thermal way but such a thermal relic is irrelevant for dark matter as its abundance turns out to be highly suppressed~\cite{Kolb:1990vq}. On the other hand, the zero mode of the axion field could be misaligned with the minimum of the axion potential whose origin is non-perturbative.  Due to this non-thermal production mechanism, the axions would be cold in the first place. In that case the axion field oscillates in its potential and one has approximately $\phi=  \phi_0 \cos(mt)$ with $m$ the axion mass close to the minimum of the axion potential. In this scenario, it is the classical energy density of the axion field that plays the role of the energy density for the cold dark matter fluid ~\cite{Arias:2012az}. In this case, the local dark matter density $\rho_0$ is related to the amplitude of the oscillation $\phi_0$ and its frequency $m$ through $\rho_0 = \frac{1}{2}m^2 \langle \phi^2 \rangle + \frac{1}{2} \langle\dot{\phi}^2\rangle = \frac{1}{2} m^2 \phi_0^2$.

It is well known that the zero-point energy of the electromagnetic field in a cavity leads to an attractive force between the walls due to the so-called Casimir pressure~\cite{Casimir:1948dh,bordag2009advances}. This is related to the fact that even if the cavity contains only vacuum, the corresponding electromagnetic modes inside the cavity are countable, as opposed to outside the cavity where they are continuous. Suppose one sets up such a cavity embedded inside a constant magnetic field. The dark matter axion field will couple to some of the modes of the vacuum and could potentially modify the value of the Casimir pressure. The question addressed here is whether this could lead to observable effects in actual experiments. Typically, there are two potentially important physical effects in this setting. The first one is classical and comes from the  force between the plates due to the induced polarisation in metals comprising the boundary plates. We discuss this effect here. The second one is a quantum effect and follows from the shift in the quantum modes of the electromagnetic field between the plaques due to the axion-photon coupling. This is left for future work, see \cite{Kharlanov:2009pv,Fukushima:2019sjn,Brevik:2021ivj,Ema:2023kvw}. We give an estimate of this effect in the appendix \ref{app:quantum}.

In this paper, we study the effect of the dark matter oscillating axion field on the Casimir pressure at the classical level. We consider a constant background magnetic field which triggers the mixing between axions and photons.
As a result an electromagnetic field is induced  along the magnetic field lines. In particular between two metallic parallel plates, we estimate the effects of this mixing on the Casimir pressure. Intuitively we expect the axion to induce resonances and thus a modification of the Casimir forces. The resonant effect on the pressure between the plates depends on the geometry of the cavity and on the frequency of the oscillating axion field. These resonances are regularized by the dissipation effect in the metal. To account for dissipation, we use a classical approach, and leave the full quantum field theoretic treatment to a future study, where the modification to the quantum pressure due to the axion field will be discussed. At the classical level, we find that the resonances are not perfect and have a finite width depending on the conductivity of the metal. The width of the resonances is also broadened by the velocity dispersion of the axion in the galactic halo. Our analysis assumes that the broadening is dominated by the metallic effects. We find that for distances of the order of a few microns, the axion-induced pressure can compete with the quantum pressure in $1/d^4$, where $d$ is the distance between the plates. We also notice that in a real material, the quantum pressure is reduced by the imperfect reflexivity of the boundary plates and we recall the Lifschitz formalism in an appendix. This would make the emergence of the axion-induced effects easier. We leave the detailed study of this to future work. The competitive cases where the axion-induced pressure becomes of the order of the quantum Casimir effect  require conductivities which are larger than the ones of ordinary metals such as copper. We leave phenomenlogical investigations on the type of metal and experimental situations, such as temperature, necessary to maximise the axionic effect to future work.

In part~\ref{equations}, we introduce the model and derive the general equations for an axion subject to a magnetic field in a geometry with boundaries.  The treatment of dissipation in metal is described there, and the classical equations of motion are obtained. In part \ref{single}, a one-sided boundary condition is considered. This is an intermediate step towards the treatment of the classical Casimir pressure induced by an axion in the presence of a magnetic field, and allows one to retrieve known results, such as the ones typically considered for dish-antenna experiments~\cite{Horns:2012jf}. Then in part~\ref{sec:cavity} we consider two boundary conditions, {\it i.e.} a cavity, with an external magnetic field parallel to the plates. We will see that in the ideal case with infinite conductivity, we obtain infinite resonances for discrete plate separations, described by Dirac distributions. We will see that the resonances are regularized when a finite conductivity is considered. In the last part we discuss orders of magnitudes for the modifications of the Casimir forces and some prospects. We discuss Green's functions in the one and two side plates, more details on dissipation, the energy budget, the Lifschitz theory \cite{lif1,lif2} and the quantum effects in the appendices.

\section{Axions in a constant magnetic field}\label{equations}
\subsection{The model}

We start with the usual Lagrangian of electrodynamics in vacuum coupled to an axion-like field $\phi$ in natural units
\be
\mathcal{L} = -\frac{1}{4} F_\munu F^\munu - \frac{\phi}{4M} F^{\munu}\tilde F_{\mu\nu},
\ee
where $\tilde F_{\mu\nu}=  \h \epsilon_{\mu\nu\rho\sigma}F^{\rho\sigma}$ is the dual field strength. We define $\epsilon^{0123}=1$. The pseudo-scalar field $\phi$  will be assumed to be time dependent, as representing the oscillations of the dark matter field in our environment. The energy scale $M$ is of the order of the scale of the Peccei-Quinn symmetry breaking. For dark matter axions of mass $m\sim 10^{-5}\;\rm eV$, this scale is at least of the order of $10^{12}\;\rm GeV$~\cite{Workman:2022ynf}. The full lagrangian includes also  kinetic terms for the axion, which we disregard here as we are interested in the conventional electromagnetic field modes. The equations of motion for the photon field become
\be
 \d_\mu F^\munu +\f{1}{2M}  \epsilon^{\mu\nu\rho\sigma}\d_\mu (\phi F_{\rho\sigma})=0.
\ee
In the Lorentz gauge  $\d_\mu A^\mu=0$, the first term gives $\Box A^\nu$ and instead of the classical d'Alembert equation we obtain the modified propagation equation
\be
\Box A^\nu  = - \f{1}{2 M} \epsilon^{\mu\nu\rho\sigma} \d_\mu (\phi F_{\rho\sigma}).
\ee
We decompose the gauge field into a background field and a perturbation
$
A^\nu  = \bar A^\nu + a^\nu
$,
where the background field is generated by a homogeneous and constant magnetic field corresponding to
$
\bar A^i = \h \epsilon^{ijk} x^j B^k,
\bar A^0 = 0,
$
satisfying the background wave equation $\Box \bar A^\nu=0$.  The Maxwell's equation for the fluctuation of the electromagnetic field  becomes
\be
\Box a^\nu = -\f{1}{2 M} \epsilon^{\mu\nu\rho\sigma}  \Big( ( \d_\mu \phi )F_{\rho\sigma} + \phi \partial_\mu F_{\rho\sigma } \Big).
\ee
In the following, we will assume that
the axion field is homogeneous and time-dependent $\d_\mu \phi = \dot{\phi} \delta_\mu^0$, corresponding to the cosmological field associated to dark matter in the environment. Then we obtain
\be
\Box a^\nu = -\f{1}{2 M} \Big( \epsilon^{0\nu\rho\sigma} \dot{\phi}\; (\bar{F}_{\rho\sigma} + f_{\rho\sigma} )
+  \epsilon^{\mu\nu\rho\sigma}\phi\; (\d_\mu \bar{F}_{\rho\sigma} + \d_\mu f_{\rho\sigma}) \Big),
\ee
and upon considering the magnetic field is constant in space,{\it i.e.} $
 \d_\mu \bar{F}_{\rho\sigma}=0
$,
we get
\be
{
\Box a^\nu = -\f{1}{2 M} \dot{\phi} \epsilon^{0\nu\rho\sigma}(\bar{F}_{\rho\sigma} + f_{\rho\sigma}) -
\f{1}{2 M} \phi\;\epsilon^{\mu\nu\rho\sigma}\d_\mu f_{\rho\sigma},
}
\ee
where last term vanishes as $f_{\mu\nu}= \partial_\mu a_\nu -\partial_\nu a_\mu$.
The two equations of motion now become
\be
{
\Box \V{a} = -\f{1}{M} \dot{\phi}\Big( \V{B} + \V{b} \Big)
 \;\;\;\;\; \text{ and } \;\;\;\;\;
\Box a^0 = 0.
}
\label{vecpot}
\ee
They are conveniently written in
terms of the perturbations of electric $e^i= -\dot a^i$ and the magnetic field $b_i= \frac{1}{2} \epsilon_{ijk} f_{jk}$ in vector form as
\beq
\Box \V{b} = -\f{1}{M} \dot{\phi} \; \vec\nabla  \wedge (\V{B} +\V{b})\\
\eeq
and
\beq
\Box\V{e}  =  \f{1}{M} \ddot{\phi} \Big( \V{B}+ \V{b} \Big) +\f{1}{M} \dot{\phi} \Big( \dot{\V{B}}+\dot{\V{b}} \Big).
\eeq
These equations can also be easily derived from a reduced Lagrangian involving an external current induced by both the background magnetic field and the oscillating axion, see appendix~\ref{sec:app_dissipation}.

\subsection{The classical field and the introduction of dissipation}

The classical equations of motions can be solved by iterations in an inverse power expansion in $M$. At leading order we have
\be
{
 \Box \V{b} = -\f{1}{M} \dot{\phi} \; \vec\nabla \wedge \V{B}
 \;\;\;\;\;  \text{ and } \;\;\;\;\;
\Box\V{e}  =  \f{1}{M} \ddot{\phi}  \V{B} +\f{1}{M} \dot{\phi} \dot{\V{B}}.
}
\label{Maxwell}
\ee
Let us denote the axion-induced source term of the electric field by
\be
\V{J} =  \f{1}{M} \ddot{\phi}  \V{B} +\f{1}{M} \dot{\phi} \dot{\V{B}},
\label{cur}
\ee
then
the perturbation of the electric field becomes
\be
\V{e}(x) = \int \D^4 u \; G_{\rm } (x;u) \V{J} (u),
\label{eq:electric}
\ee
in terms of a well-defined  Green's function which depends on the geometry of the experimental situation under consideration. Here $x^\mu$ and $u^\mu$ are 4-vectors. The Maxwell equation (\ref{Maxwell}) for the electric  with the current (\ref{cur}) characteristic of axion electrodynamics have important consequences. First of all, let us notice that the current $\vec J$ is parallel to the external magnetic field. This implies that the electric field in(\ref{Maxwell}) sourced by the external current is parallel to the external magnetic. In fact the Maxwell equation (\ref{vecpot}) for the vector potentials shows that the classical sourced solution is also along the external magnetic field.  

In the cases of interest, we will consider experimental situations where the background magnetic field is present everywhere in space. The boundary conditions used to exhibit the{classical Casimir effect correspond to a cavity between two parallel plates where the external magnetic field is along the plates in the  $x$ direction when the direction perpendicular to the planes is chosen to be the $z$ direction. As we are interested in the classical solutions sourced by the external current $\vec J$ along the $x$ direction, the gauge field $\vec a$ solution to (\ref{vecpot}) only depends on $z$ direction by symmetry reason. A dependence of the solution on the $(x,y)$ directions would break the translation invariance of the cavity configuration with its constant magnetic field parallel to the plates\footnote{Solutions of the vacuum Maxwell equation in the cavity can propagate along the plane. Although they are crucial for the calculation of the quantum Casimir effect, they are not generated by the external source here.}. As a consequence the Maxwell equation (\ref{vecpot}) becomes a scalar equation for a single component of the gauge potential along the $x$ direction with a dependence on time and the $z$ direction. Hence the Green's function becomes a scalar Green's function $G$ instead of the tensor Green's function which can be found in \cite{Landau9}\footnote{The off-diagonal terms in the tensor Green's functions can be seen to vanish when no dependence on $(x,y)$ exists. This can be seen explicitly in eq. (81.5) of chapter 81 in \cite{Landau9}  dedicated to quantum effects in cavities. This results from taking $q=0$  in eq. (81.5). }. As the induced vector potential is along the external magnetic field, so is the induced electric field. Moreover as the vector potential only depends on $z$, the induced magnetic field is along the $y$ direction and therefore parallel to the plate too.  Inside the plates, the perturbations of the electric and magnetic field penetrate only over a length depending on  $\omega_{\rm Pl}^{-1}$, {\it i.e.} the skin depth $\delta$.

Before moving to the estimates of the classical pressure in specific cases, we must discuss the treatment of dissipation inside matter. This is relevant for the case where the metallic plates are made of non-ideal conductors. For this we follow the method described in~\cite{landau}.
Inside matter as in the case of metals, the relevant vector field for electric phenomena is the displacement field $\vec d= \vec e + \vec p$, sometimes called the electric induction field. It differs from the electric field in vacuum due to the existence of the polarisation $\vec p$. The displacement vector is related to the electric field via a retarded effect
\be
\vec d(\vec x, t) = \int_{-\infty}^{+\infty} d\tau \epsilon(t-\tau) \vec e(\vec x, \tau)\equiv \epsilon\star_t \vec e(\vec x,t),
\ee
or equivalently $\vec d(\vec x, \omega)= \epsilon(\omega) \vec e(\vec x, \omega)$
in Fourier space with
\be
\vec d(\vec x, t)= \int \slashed{d} \omega \epsilon (\omega) \vec e(\vec x, \omega)e^{-i\omega t},\label{eq:displacement}
\ee
with $\slashed{d} = d/2\pi$. Dissipation will be taken into account considering that the permittivity is not equal to one, and is a complex number. This is because the propagation equation in matter in the $z$ direction is $\partial^2_z e = \epsilon \partial^2_t e $ and if $\epsilon \in \mathbb{C}$, then from the dispersion relation $k^2 = \epsilon \omega^2$, $k$ is also a complex number $k = k' + i k''$, and a plane wave gets a real factor that can account for dissipation $e^{-k'' z} e^{-i(\omega t- k' z)}$.

In the following we are interested in the case of a metal, and we follow the Drude model. Readers familiar with this model and only interested in the applications to axion physics can skip this paragraph. The permittivity relates the polarisation field to the electric field through $\vec{p} = (\epsilon-1) \vec{e}$. On the one hand one has in the $x$ direction $p=Nq x$, where $N$ is the density of electrons and $q$ their charge. The movement of electrons are modeled classically with a frictional force and a restoring force, such that
\be
\ddot{\vec{x}} = \frac{q}{m}\left ( \vec{e} +  \dot{\vec{x}}\wedge (\vec{B}+\vec {b})\right ) - \gamma  \dot{\vec{x}} - \omega_0^2 \vec{x},
\ee
where $\gamma$ is a typical damping time and $\omega_0^2$ is related to the confinement of electrons around atoms. Typically we expect that $\gamma = N v_e \sigma_{\rm int}$ where $\sigma_{\rm int}$ is the cross section of the moving electrons with the material and $v_e= \sqrt{\frac{3T}{m_e}}$ is the thermal velocity of the electrons. The Lorentz force acting on the electrons depend on both the external magnetic field $\vec B$ and the induced one $\vec b$. We take the external magnetic field in the $x$ direction and we consider the induced electric waves as propagating in the $z$ direction by symmetry reason. Because the electric field perturbation is in the direction of the external magnetic field, the only displacement of the electrons related to $\vec{e}$ is in the $x$ direction and as far as the  polarisation is concerned, the magnetic force has no effect. Indeed, the external magnetic field is along the motion of the electrons and as usual we neglect the force coming from the induced magnetic since its magnetic is reduced compared to the electric force by a factor of $v/c\ll 1$. Solving the equation of motion in Fourier space one gets
\be
x=\frac{q/m}{\omega_0^2 - \omega^2 - i\gamma \omega} e.
\ee
In metals, electrons are free and $\omega_0\rightarrow 0$ so it is disregarded. By identification with the expression $\vec d= \vec e + \vec p$, one gets for the complex permittivity
\be
\epsilon= 1- \frac{\omega_{\rm Pl}^2}{\omega^2 +i\gamma  \omega },\label{eq:epsilon}
\ee
where $\omega_{\rm Pl} = \sqrt{Nq^2/m}$ is the plasma frequency.  The damping time $\gamma$ is related to the conductivity through
\be
\sigma= \frac{\omega^2_{\rm Pl}}{\gamma}.
\ee
The integral leading to the displacement field $\vec {d} (\vec {x},t)$ can be evaluated in the complex plane. In the Drude model, when $t<0$ the contour must be closed in the upper half plane and as long as $\vec {e} (\vec {x},\omega)$ has no singularity in the upper half plane  we find that $\vec d(\vec x, t)=\vec e (\vec x,t)$. On the contrary, when $t\ge 0$, the contour must be closed in the lower half plane where the permittivity has a pole at $\omega_p= -i\gamma$. Noticing that for the Drude model  $\vert \omega \vert \vert \epsilon (\vec x, \omega)-1 \vert$  converges to zero for large  $\vert \omega\vert $, the integral on the large circle in the lower half plane vanishes and  we find that
\be
\vec{d} (\vec x, t)= \vec e (\vec x,t) +  \sigma  \vec e(\vec x, -i\gamma)e^{-\gamma t},
\ee
implying that the displacement field vanishes after a time $1/\gamma$ and has an amplitude depending on the conductivity. This is what happens due to dissipation as long as a permanent regime is not considered.

Including the displacement field, the phenomenological equation for the electric field is
\be
-\partial_0 \left ( \epsilon \star_t \partial_0 G \right ) + \Delta G = \delta^{(4)}(x^\mu- y^\mu).
\label{eq:green}
\ee
This equation can be solved using the Green's function of the operator in Fourier space
\be
G\equiv (\Delta + \epsilon(\omega) \omega^2)^{-1}.
\ee
The Green's function can be explicitly defined as soon as the boundary conditions are specified. In the next parts of the paper, different boundary conditions are considered, and the associated Green's functions are used to solve the equations of motions in the vacuum and in the metallic plates.

\subsection{Classical pressure from the electromagnetic field and the axion field}

In Casimir experiments, the observable is an attractive force between the plates. This can be described by a negative pressure in vacuum. Before moving on with the expression of the pressure, several remarks are in order. First of all, an interesting effect follows from the Maxwell equation in matter.
Inside the plaques we have the identity
\be
\vec \nabla \wedge \vec b = \partial_0 \vec e + \vec j_{\rm ind},
\ee
where $\vec j_{\rm ind}$ is the induced current. As $\vec\nabla \wedge \vec b - \partial_0 \vec e= -\Delta \vec a +\partial_0^2 a$, we find that
$\vec j_{\rm ind}= -\partial_0 ( \epsilon \star_t \partial_0 \vec a - \partial_0 \vec a)$ which coincides with
\be
\vec j_{\rm ind}= \dot {\vec p}
\ee
inside matter. As can be seen this current only exists in a finite width within the plate, {\it i.e.} this is a skin effect. This current is responsible for the dissipation power $\vec j_{\rm ind}. \vec e$ in the plate, corresponding to the loss of energy $\frac{d}{dt}(\frac{\vec p.\vec e}{2})$ due to the
polarisability of the material. As a result, the plaque heats up due to the Joule effect. We will give a thorough discussion of dissipative effects in appendix \ref{ap:budget}.

The oscillating axion field will perturb the electromagnetic modes via the coupling between the axion and two photons. In Casimir experiments, the observed effect is a pressure in vacuum. Here there is an additional classical pressure that will be induced by the coupling of the axion field with the electromagnetic field.
In vacuum, the energy momentum tensor of the electromagnetic field  is given by
\be
T^{\mu\nu}= F^{\mu\rho}F^\nu_\rho -\frac{\eta^{\mu\nu}}{4} F^2,
\ee
which gives for the pressure against the plates
\be
\sigma^v_{zz}=- T^{zz}= \frac{e_z^2+b_z^2}{2} -\frac{e^2+b^2}{2}.
\ee
Notice that there is no contribution from the axion term as $F\tilde F$ is topological and independent of the metric. Indeed $T^{\mu\nu}$ is obtained by the variation of the action with respect to the metric; the $F\tilde{F}$ term gets a $(-g)^{-1/2}$ term for the Levi-Civita tensor to be covariant, that cancels with the $\sqrt{-g}$ in the integration measure of the action. Moreover $e_z=b_z=0$ implying that the electromagnetic pressure on the plate due to the axion source
is given by
\be
\sigma^v_{zz}= -\frac{e^2+b^2}{2},
\ee
evaluated on the plates. This is the pressure due to the vacuum on the interface between the plaques and the vacuum.

In matter, the energy-momentum tensor of electromagnetism gives the pressure
\be
\sigma^m_{zz}= \frac{e_zd_z+ b_z^2}{2} -\frac{\vec e. \vec d+b^2}{2} = -\frac{\vec e. \vec d+b^2}{2} .
\ee
As $e_z=0$, the pressure on the plaques is given by the difference from the two classical pressures inside and outside the plaques. This gives
\begin{widetext}
\be
P= \sigma_{zz}^v- \sigma_{zz}^m = -\frac{e^2}{2} + \frac{\vec{e}\cdot\vec{d}}{2} = - \frac{1}{2} \vec{e}\cdot \left (\vec{e} - \vec{d} \right ) = \frac{\vec p.\vec e}{2}  = \Re\left ((\epsilon-1) e^2 \right )\label{eq:pressure}.
\ee
\end{widetext}
This term is always negative.
In the next sections, the explicit determination of the electric field and the related polarisation field,  allows for the computation of the pressure. In the single-plate case, this leads to a thrust force on the plate, whereas in the two-plate case, this modifies the Casimir effect.

\section{Radiation from a single plate}
\label{single}

As an application of the Green's function techniques, we will derive the expression for the electric field when only one plate is present and the magnetic field penetrates inside a metal with finite conductivity. This generalizes the ideal case where the conductivity is taken to be infinite. The geometry for this section is sketched in Fig.~\ref{fig:oneplate}.
\begin{figure}[h]
\centering
\includegraphics[width=\columnwidth]{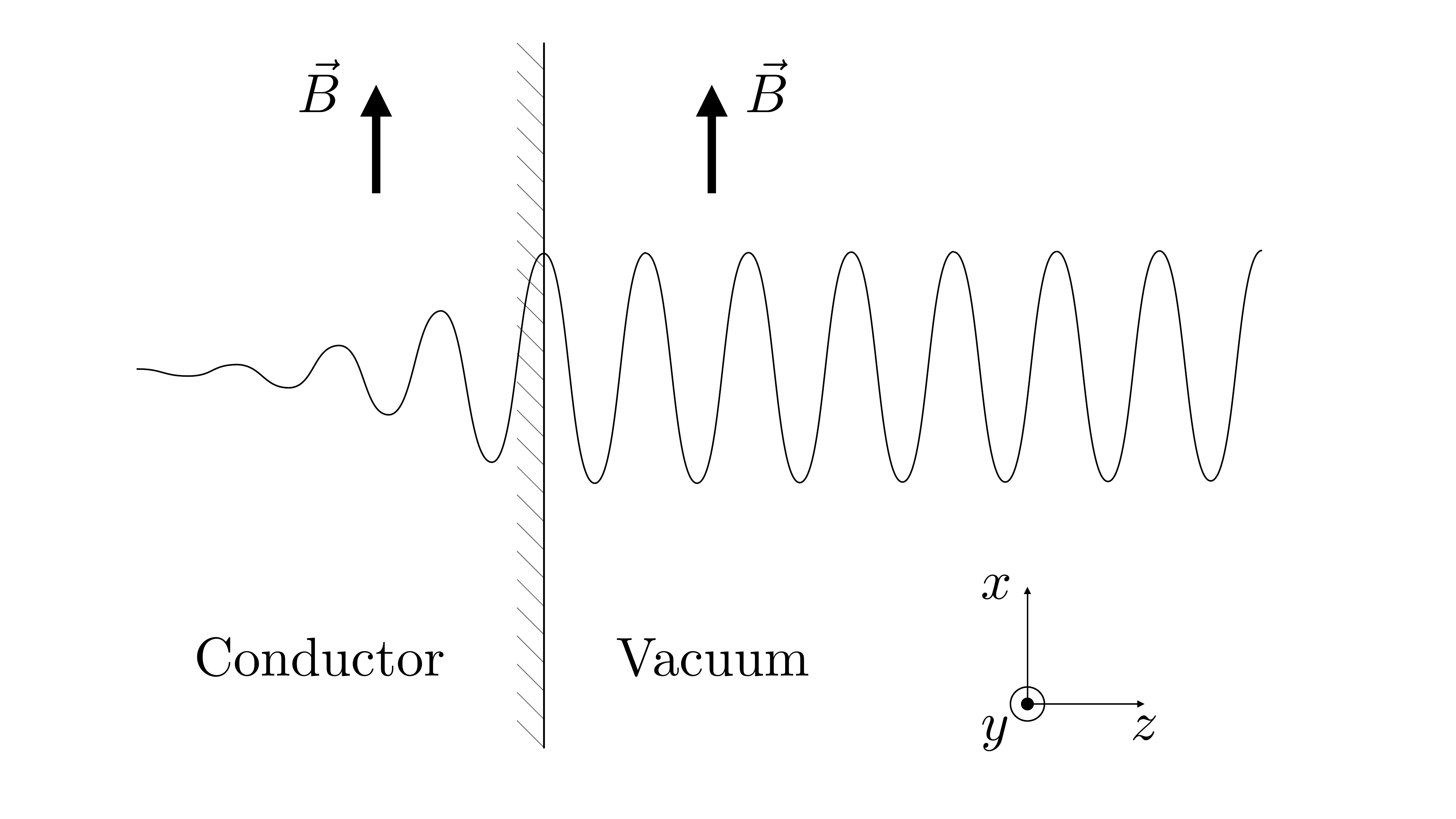}
\caption{Geometry and boundary conditions for the one-plate case.}
\label{fig:oneplate}
\end{figure}

Now we consider the generation of an electric field due to the axion coupling when the magnetic occupies all space. This is deduced from Eq.~\ref{eq:electric} as a function of the Fourier transform in time of the Green's function evaluated at the frequency $\omega=m$,
\be
e(z,t)= J_0 \Re \left [ e^{-im t} \int_{-\infty}^\infty dz_0  G(m, z;z_0) \right ],
\ee
where  in the case of an external magnetic field $B_0$ and a homogeneous axion field, the source term  depends on the amplitude
$$
J_0 = \frac{m^2\phi_0 B_0}{M}.
$$
The Green's function is such that it vanishes in vacuum at $z=+\infty$ and in matter at $z=-\infty$.
In vacuum, this selects the modes in $e^{i\omega z}$ where the replacement $\omega^2 \to \omega^2 +i\tilde \epsilon$ with $\tilde \epsilon \to 0$ guarantees the convergence. The Green's function is solution of Eq.~\ref{eq:green}. To determine explicitly $G$, it is convenient to separate the cases $z_0<0$ and $z_0>0$, which is done in the next subsections. More details on the derivation of the Green's function will be given in the case of two-plate systems.

\subsubsection{Green's function for $z_0<0$}
We find that when $z<z_0$,
\be
G(\omega,z;z_0)= \frac{1}{2\xi(\omega)} \left (1+ \frac{\xi(\omega) +i\omega}{\xi(\omega)-i\omega}e^{2\xi(\omega)z_0}\right ) e^{\xi(\omega)(z-z_0)},
\ee
whilst when $z\in [z_0,0]$,
\be
G(\omega,z;z_0)= \frac{e^{\xi(\omega)z_0}}{2\xi(\omega)}\left (e^{-\xi(\omega) z}+\frac{\xi(\omega)+i\omega}{\xi(\omega)-i\omega}e^{\xi(\omega)z}\right ),
\ee
and finally when $z>0$,
\be
G(\omega,z;z_0)= \frac{e^{i\omega z +\xi(\omega) z_0}}{\xi(\omega) -i\omega},
\label{eq:G1}
\ee
where we defined the function $\xi(\omega)$ as
\be
\xi(w) = \left ( - \epsilon(w) \omega^2 \right )^{1/2},\label{eq:xi}
\ee
where the square root is such that the real part is always positive. The same calculation can be performed when $z_0>0$.

\subsubsection{Green's function for $z_0>0$}
In this case, it is shown in  appendix~\ref{ap:single} that if $z<0$,
\be
G(\omega,z;z_0)= \frac{e^{i\omega z_0+ \kappa(\omega) z}}{i\omega +\xi(\omega)},
\ee
if $z\in [0,z_0]$,
\be
G(\omega,z;z_0)=\left ( \frac{1}{i\omega +\xi(\omega)}-\frac{1}{2i\omega}\right )e^{i\omega (z+z_0)}+ \frac{e^{i\omega (z_0-z)}}{2i\omega},\label{eq:G2}
\ee
and finally when $z>z_0$,
\be
G(\omega,z;z_0)= \left ( \frac{1}{i\omega +\xi(\omega)}-\frac{1}{2i\omega}\right )e^{i\omega (z+z_0)}+ \frac{e^{i\omega (z-z_0)}}{2i\omega}.\label{eq:G3}
\ee

\subsubsection{The electric field in vacuum}

In dish antenna experiments, the boundary condition implied by the presence of the conductor leads to the emergence of an outgoing electric field wave. This propagating electric field is the signal of interest.
To compute its amplitude, we now have to select the Green's function for $z>0$, i.e. outside the plate in equations~\ref{eq:G1},~\ref{eq:G2} and~\ref{eq:G3}. As a result we have
\begin{widetext}
\be
\int_{-\infty}^\infty dz_0  G(m, z;z_0)= \frac{1}{m^2} + \left ( \frac{1}{\xi(m)(\xi(m)-im)}- \frac{\xi(m)}{m^2}\frac{1}{\xi(m)+im}\right )e^{imz}.
\ee
\end{widetext}
Hence there are two components. There is an oscillating electric field at the frequency $m$ of the axion oscillations. This is due to the presence of the magnetic field in the vacuum and the $\phi\vec{E}\cdot\vec{B}$ form of the axion coupling : as soon as a magnetic field line is present, a small electric field is induced. There is also a propagating wave with frequency $m$, due to the presence of the boundary condition at $z=0$ and the breaking of translation invariance in that direction. Explicitly written, the electric field reads
\begin{widetext}
\be
e(z,t)= \frac{J_0 \cos mt}{m^2}+ J_0 \Re \left [ \left ( \frac{1}{\xi(m)(\xi(m)-im)}- \frac{\xi(m)}{m^2}\frac{1}{\xi(m)+im}\right)e^{im(z-t)}\right ].
\ee
\end{widetext}
It is interesting to take the limit of an ideal metal corresponding to $\xi(m)\to \infty $. In this case we find that the electric field becomes
\be
e_{\rm ideal}(z,t)=\frac{J_0}{m^2}\Re  \left [ e^{-imt}(1-e^{imz})\right ].
\ee
As expected the oscillating solution is simply
\be
e_{\rm osc}(t)= \frac{J_0 \cos mt}{m^2},
\ee
directly from the propagation equation in the absence of $z$ dependence. Mathematically, this is the solution of the propagation equation including the source term. The propagating wave satisfies the propagation equation in the absence of the source term and its amplitude is such that the electric field vanishes at the surface of the perfect conductor. This specifies a unique solution. Here this has been obtained using the Green's function and has been generalized to the case of a non-perfect conductor.
The other component is a propagating electric field in the $z$ direction,
\be
e(z,t)_{\rm prop} = \frac{J_0}{m^2} \cos \left ( m(z-t) \right ).
\ee
This expression allows us to retrieve the usual expressions for dish antenna experiments. The output power per unit area is given by the Poynting vector
$$
\Pi = \langle e^2 \rangle = \frac{\rho_0 B_0^2}{ m^2 M^2}
$$
where the average is taken over the rapid oscillations. With conventional units and some typical values for the parameters, one finds

\begin{widetext}
\be
\Pi_{\rm ideal} = 2.76\times 10^{-30}{\;\rm W/m^2}
\left ( \frac{\rho_0}{0.3 \rm \; GeV/cm^3} \right)
\left ( \frac{B_0}{ 1 \;\rm T} \right)^{2}
\left ( \frac{m}{100\;\rm \mu eV} \right)^{-2}
\left ( \frac{M}{10^{14}\;\rm GeV} \right)^{-2}.
\ee
\end{widetext}

The previous expression is used to estimate the signal power in dish antenna experiments in the case a perfect conductor setting the boundary condition. It is interesting to see how the Green's function method developed here allows for a  generalisation  to the case of non-ideal metals.  To obtain a useful formula, one can consider for real metals that $\omega\ll \gamma \ll \omega_{Pl}$, in which case
\be
\Pi_{\rm real} = \frac{\rho_0 B_0^2}{ M^2} \left (\frac{1}{m^2} + \frac{1}{\sigma^2} \right ) = \Pi_{\rm ideal}  + \Pi_{\sigma}.
\ee
This leads to a tiny correction of the emitted power, which is at first order independent of the axion mass,
\begin{widetext}
\be
\Pi_{\sigma} = 4.55\times 10^{-35}{\;\rm W/m^2}
\left ( \frac{\rho_0}{0.3 \rm \; GeV/cm^3} \right)
\left ( \frac{B_0}{ 1 \;\rm T} \right)^{2}
\left ( \frac{\sigma}{6\times 10^8\;\rm /\Omega/m} \right)^{-2}
\left ( \frac{M}{10^{14}\;\rm GeV} \right)^{-2},
\ee
\end{widetext}
where the reference value for the conductivity is that of Copper. Notice that a finite conductivity, and thus dissipation leads to an increase of the signal power. This is rather counter-intuitive, but allowing for $\vec{p}\neq\vec{0}$ leads to more small oscillating dipoles, hence more radiation. As can be seen from the previous formula, the effect is very small.

\subsubsection{Pressure on a single plate}

The emission of an electromagnetic wave from the surface of the plate extracts momentum from the electromagnetic field, thus leading to a force per unit area. This is simply the radiation pressure, and is due to the presence of the polarization field at the surface of the conductor. The pressure is given by a time average of Eq.~\ref{eq:pressure},
\be
P= \frac{\langle \vec e(0,t). \vec p(0,t)\rangle}{2} \label{eq:pressureaverage}
\ee
where $\vec p$ is the polarisation vector, which is deduced from Eq.~\ref{eq:displacement},
\be
\vec p(z,t)= \int \slashed{d} \omega (\epsilon(\omega)-1) \vec e(z,\omega).
\ee
Using the results from the previous section, we have that $\vec p$ is along $\vec B$ with a magnitude
\begin{widetext}
\be
p(0,t)= J_0 \Re  \big ( (\epsilon (m)-1)\left ( \frac{1}{m^2} +  \frac{1}{\xi(m)(\xi(m)-im)}- \frac{\xi(m)}{m^2}\frac{1}{\xi(m)+im}\right) e^{-imt}\big ).
\ee
\end{widetext}
So in that case, the averaged pressure on the plaque is given by
\begin{widetext}
\be
  P  = \frac{J_0^2}{2} \Re \left [ \epsilon(m) -1 \right ] \left \vert \frac{1}{m^2}+  \frac{1}{\xi(m)(\xi(m)-im)}- \frac{\xi(m)}{m^2}\frac{1}{\xi(m)+im} \right \vert ^2
\ee
\end{widetext}
which is always negative, {\it i.e.} the vacuum attracts the plaque.
This pressure is rather small for typical values of the parameters. For instance, for a 50~T magnetic field, an axion mass of $m=100 \;\rm \mu eV$ and a coupling scale $M = 10^{14}\;\rm GeV$, the value of the pressure is of the order of $10^{-39}\;\rm Pa$. One can imagine using this pressure to produce thrust on a spaceship.
Should a spaceship be equipped with such a system and a 1~$\rm m^2$ metal plate, the pressure converts into a $10^{-40} \; \rm kg$ thrust. As a comparison, the same plate at the level of the Oort cloud would undergo a thrust of about $10^{-17}\;\rm kg$ from Solar radiation pressure. { On the other hand as dark matter is present everywhere even in the absence of radiation from nearby stars, we could imagine that such an effect could be used in total darkness}.

\section{Classical pressure in an empty cavity}
\label{sec:cavity}

\subsection{The ideal case}

Let us move on to  the simple case of a two-plate system, where the plates are made of an ideal conductor, {\it i.e.} with infinite conductivity. In that case the electric field do not penetrate the metallic plates, as sketched in Fig.~\ref{fig:twoplatesideal}.

\begin{figure}[h]
\centering
\includegraphics[width=\columnwidth]{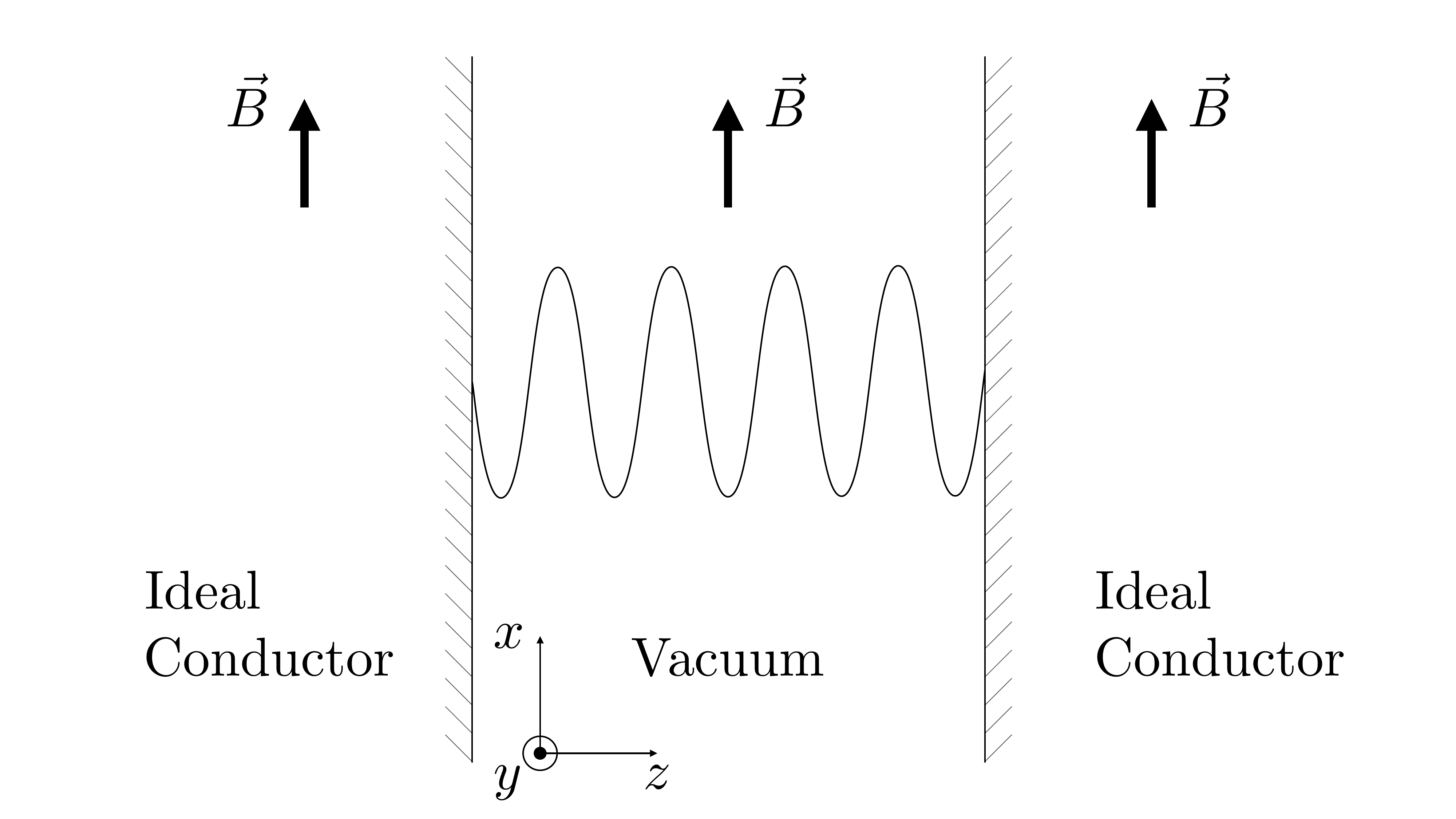}
\caption{Geometry and boundary conditions for the two-plate case with perfect conductors.}
\label{fig:twoplatesideal}
\end{figure}

Like in the previous situation, the  electric field $\vec e $ points in the $\vec B$ direction. We denote by $e=\frac{\vec e. \vec B}{B}$ its magnitude such that $\Box e= J$. Denoting by $G(z,t;z_0,t_0)$ the Green's function such that
\be
-\ddot G + G''= \delta(t-t_0) \delta (z-z_0),
\ee
where $'=d/dz$ we have then
\be
e(z,t)= \int dz_0 dt_0 G(z,t;z_0,t_0) J(z_0,t_0).\label{eq:electricideal}
\ee
We now perform the computation of the Green's function in that case.

\subsubsection{Determination of the Green's function}

Using Fourier modes in time, the Green's function verifies
\be
G'' + \omega^2 G= \delta (z-z_0) e^{i\omega t_0}.
\ee
Denoting by $d$ the distance between the plates in the $z$ direction, the solution becomes
\begin{eqnarray}
&&0\le z\le z_0, \ \  G= A \sin \omega z
\nonumber \\
&&
z_0\le z\le d, \ \ G= C \sin \omega z + D\cos \omega z.
\nonumber
\end{eqnarray}
We impose the Dirichlet boundary conditions, that is $e(0,\omega)=e(d,\omega)=0$. This corresponds to the non-penetration of the electric field inside the plates and the continuity of its parallel component at the interface. This implies that
\be
C \sin \omega d + D \cos \omega d=0.
\ee
We must also have a discontinuity of $G'$ with a jump
\be
[G']_{z_0}= e^{i\omega t_0}.
\ee
We find that
\be
C=\frac{\sin \omega z_0}{\omega} \frac{e^{i\omega z_0}}{\tan \omega d}, \ \ D= -\frac{\sin \omega z_0}{\omega}{e^{i\omega z_0}}
\ee
and
\be
A= \frac{\sin \omega z_0}{\omega} \frac{e^{i\omega z_0}}{\tan \omega d}\left (1- \frac{\tan \omega d}{\tan \omega z_0}\right ).
\ee
Now we have
\be
J= \frac{m^2 B\phi_0}{M} \cos m t \equiv J_0 \frac{e^{imt}+e^{-imt}}{2}.
\ee
Using
\be
 G(z,t;z_0,t_0)= \int \slashed{d}\omega e^{-i\omega t} G(z,\omega; z_0,t_0),
\ee
and
\begin{widetext}
\ba
&& 0\le z\le z_0 \ \   G(z,\omega; z_0,t_0)=\frac{e^{i\omega t_0}}{\omega}\left ( \frac{\sin \omega z_0}{\tan \omega d}- \cos \omega z_0\right ) \sin \omega z \nonumber \\
&& z_0 \le z\le d \ \ G(z,\omega; z_0,t_0)=\sin \omega z_0 \frac{e^{i\omega t_0}}{\omega}\left (\frac{\sin \omega z}{\tan \omega d}- \cos \omega z\right ) \nonumber \\
\ea
\end{widetext}
we first evaluate
\be
I_1= \int \slashed{d}\omega \frac{e^{i\omega (t_0-t)}}{\omega}\frac{\sin \omega z \sin \omega z_0}{\tan \omega d}.
\ee
The integrand is a meromorphic function in the complex plane with no singularity at the origin and simple poles for
\be
\omega_n= \frac{n\pi}{d} \text{, with }  n\in \mathbb{N}^\star.
\ee
If $t>t_0$, we close the contour on a large circle in the lower half plane, avoiding all the poles with infinitesimal circles above the horizontal axis
\be
I_1= - \pi i \sum_{n\ne 0} \frac{\rm Res_n}{2\pi},
\ee
where ${\rm Res_n}= \frac{e^{i\omega_n(t_0-t)}}{d\omega_n} \sin \omega_n z \sin \omega_n z_0$. The same method holds when $t<t_0$ by closing the contour in the upper half plane.
For the same reason, by going to the complex plane we have
\be
\int \slashed{d}\omega \frac{e^{i\omega (t_0-t)}}{\omega}\sin \omega z \cos \omega z_0 =0,
\ee
leading to
\be
G(z,t;z_0,t_0)=\sum_{n>0} \frac{\sin \omega_n z\sin \omega_n z_0}{d\omega_n} \sin \omega_n (t_0-t).
\ee
We see that the Green's function oscillate at the resonance frequencies $\omega_n$.

\subsubsection{Resonances}
The final step for this evaluation of the electric field  consists in performing the time integral in Eq.~\ref{eq:electricideal}, first we obtain
\be
e(z,t)= - \pi J_0 \sum_{n>0} \frac{\sin \omega_n t \sin \omega_n z}{d\omega_n^2}(1- \cos \omega_n d) \delta (\omega_n -m).
\ee
Notice that the sum vanishes for all even n, implying that
\be
e(z,t)= - 2\pi J_0 \sum_{p\ge 0} \frac{\sin \omega_{2p+1} t \sin \omega_{2p+1} z}{d\omega_{2p+1}^2} \delta (\omega_{2p+1} -m).
\ee
The delta function selects one specific frequency, at which there is a resonance when $\omega_{2p+1}=m$. This specifies a number of distances $d_p= (2p+1)\pi/m$  where the electric field is resonant. The energy and pressure associated with these resonances is ill defined as it involves the square of Dirac distributions. This resonance is regularized when one takes into account the finite conductivity of the plates. In particular, we will find that the finite conductivity leads to a shift of the resonances below the real axis in the complex plane of pulsations $\omega$. Moreover, the number of resonances will become finite.

\subsection{Non-ideal conductors with no magnetic field inside}

As a generalization of the previous situation, and in order to regularise the resonances obtained before, we now perform the same study with the inclusion of a finite conductivity for the metallic plates. For pedagogical purpose, the study is performed in two steps, first one considers no magnetic field penetrates the metallic plates. This would correspond to superconducting material and leads to simpler expressions. Then, the more realistic case of a magnetic field present everywhere, including in the plates, is considered.

\subsubsection{The Green's function}

We  now introduce a  finite conductivity for the boundary metallic plates, and consider the magnetic field only lies outside the plates. It corresponds to the situation sketched in Fig.~\ref{fig:twoplatessuperc}. In the next section, we will extend the setting to the more realistic case where both the conductivity is finite and the magnetic field penetrates inside the metal.

\begin{figure}[h]
\centering
\includegraphics[width=\columnwidth]{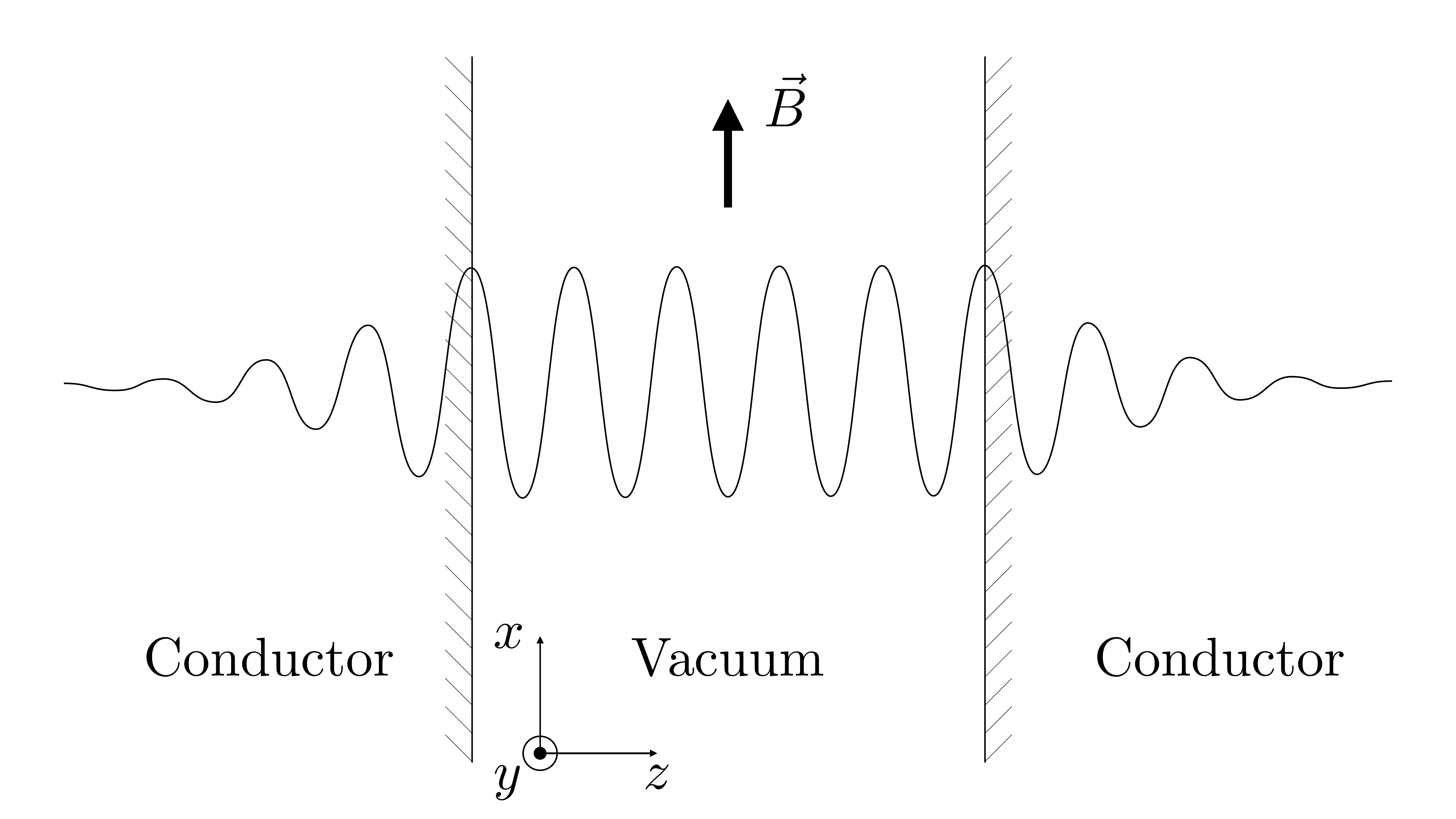}
\caption{Geometry and boundary conditions for the two-plate case with finite-conductivity but no magnetic field.}
\label{fig:twoplatessuperc}
\end{figure}

Inside the plates the permittivity is not equal to one and is given by~\ref{eq:epsilon}. Notice that there is a pole in the lower half plane as required by causality, {\it i.e.} the support of the Fourier transform of the permittivity $\epsilon (t)$ as a function of time is the positive real axis implying that the displacement field at a given time depends only on the values of the electric field in the past.
In this case, the Green's function satisfies
\be
 G'' + \epsilon(\omega) \omega^2 G= e^{i\omega t_0} \delta (z-z_0)
\ee
 for $z<0$ and $z>d$, together with
\be
 G'' + \omega^2  G= e^{i\omega t_0} \delta (z-z_0)
\ee
for $0\le z\le d$. The values of the Green's function for different ranges in $z$ are shown in Tab.~\ref{tab:green}. The values for $G_+$ and $G_-$ are specified below.

\begin{table}[h]
\centering
\begin{tabular}{c|c}
Range for $z$ & Expression of the Green's function \\ \hline \hline
$z<0$ & $G=G_- e^{\xi z},\ z<0$ \\ \hline
$0\le z\le z_0$ & $G= A \sin \omega z + B \cos \omega z$ \\ \hline
$z_0\le z\le d$ & $G= C \sin \omega z + D\cos \omega z$ \\ \hline
$z>d$ & $G= G_+ e^{-\xi(z-d)}$
\end{tabular}
\caption{Expressions of the Green's function in different $z$ ranges.} \label{tab:green}
\end{table}

Notice that here $z_0$ lies between $0$ and $d$. The general case is treated in appendix \ref{app:G}.
We find that
\be
G_-= \Theta (\omega) {e^{i\omega t_0}}\left ( \cos \omega (z_0-d) - \frac{\xi}{\omega} \sin \omega(z_0-d)\right )
\ee
and
\be
G_+= \frac{\Theta (\omega)}{\sin \omega d} {e^{i\omega t_0}}\left ( \cos \omega z_0 + \frac{\xi}{\omega} \sin \omega z_0\right ),
\ee
where $\Theta(\omega)$ is given by
\be
\Theta (\omega)= \frac{1}{\omega\left ( (1- \frac{\xi^2}{\omega^2}) \sin \omega d -2 \frac{\xi}{\omega} \cos \omega d\right )}.\label{eq:theta}
\ee
With this, we can identify
\begin{widetext}
\ba
&& 0\le z\le z_0, \ G= G_- \left ( \cos \omega z +\frac{\xi}{\omega} \sin \omega z\right )\nonumber \\
&& z_0 \le z \le d, \ G= G_+\left ( \cos \omega(z-d) -\frac{\xi}{\omega} \sin \omega (z-d)\right) .
\ea
\end{widetext}
This Green's function depends on the frequency $\omega$ and has resonances when $\Theta (\omega)$ has poles. The nature of the resonances is important as will be seen below.

\subsubsection{The resonances}

The function $\Theta$ has poles
 when $\omega$ verifies
\be
\tan \omega d= \frac{2\frac{\xi}{\omega}}{1- \frac{\xi^2}{\omega^2}}.
\label{eq:pole}
\ee
We can distinguish several cases.

When $\omega \gg \omega_{\rm Pl}$, the resonances would be for
\be
\tan \omega d \simeq 2i,
\ee
which has no solution. In this limit the Green's function is
\ba
&& z\le z_0\ G \simeq \frac{e^{i\omega( t_0 +z_0-z)}}{2i \omega}\nonumber \\
&& z\ge z_0\ G\simeq \frac{e^{i\omega( t_0 +z-z_0)}}{2i \omega},
\ea
with no effect from the plates at all, {\it i.e.} this is the free propagator in (1+1) dimensions and the plates are transparent at high frequency. This is the optical regime
of a metal as one expects for frequencies larger than the plasma frequency.

Below the plasma frequency, there is another characteristic scale given by $\gamma$.
The poles are now obtained as solutions of
\be
\tan \omega d \simeq -\frac{2}{(-\epsilon)^{1/2}},
\ee
which gives
\be
\omega_n d \simeq n\pi \left (1- \frac{2}{d\omega_{\rm Pl}}\right ) - \frac{i\gamma}{\omega_{\rm Pl}}
\label{reso}
\ee
as long as $\omega_n \gg \gamma$ .
When $\omega \ll \gamma$, the poles are located at
\be
\omega_n d\simeq n\pi -2 e^{i\pi/4} \frac{\sqrt{n\pi \gamma}}{\omega_{\rm Pl} \sqrt d}.
\ee
Resonances exist for $\omega_n \ll \gamma$ only if $d\gg \gamma^{-1}$.

As a result all the poles are shifted by the effect of the plasma frequency and are now below the real axis.  Notice that there are only a finite number of poles below the plasma frequency, whereas there were an  infinite number in the ideal case. Poles below the real axis imply that the Green's function respects causality.
In fact, the poles of $\Theta$ represent the eigenfrequencies of the free system with no source
\be
 a'' +  \epsilon(\omega)\omega^2 a= 0.
\ee
These modes have a temporal dependence in $e^{-i\omega_n t}$ which goes to zero at large time, {\it i.e.} the modes are evanescent waves which decay due to the dissipation induced by a finite conductivity.

\subsubsection{The electric field}

We can now evaluate the electric field with metallic plates on the boundary.
Formally the electric field is given by
\be
e(z,t)= J_0 \Re \left [ \int_0^d dz_0 \int \slashed{d} \omega e^{-i\omega t} \slashed{\delta}( \omega -m) G(\omega, z;z_0, 0) \right ],
\ee
where $G(\omega, z;z_0, t_0)$ was calculated in the previous section. Thanks to time translation invariance we can always set $t_0=0$ and use the expressions derived in the appendix \ref{app:G} obtained by setting $p_\parallel=0$ as we consider that the problem is planar.
Notice that the integral on $z_0$ is only between the plates as the magnetic field vanishes in the plates. Now the integral over $\omega$ is trivial if $G$ has no singularities along the real axis. We have seen that there are only poles when $\Theta$ diverges. This can happen
when~\ref{eq:pole} is satisfied, {\it i.e.} for a finite number values for $\omega$, or when $\omega=0$. Close to $\omega=0$, we have $\Theta \simeq -1/(2\omega^2_{\rm Pl}d+ \omega_{\rm Pl})$ so no singularity. As a result when the plasma frequency does not vanish, the Green's function has no singularity on the real axis and therefore
\be
e(z,t)= J_0 \Re \left [ e^{-im t} \int_0^d dz_0  G(m, z;z_0, 0) \right ].
\ee
As a result, we have
\begin{widetext}
\be
z\le 0, \ \ e(z,t)= J_0 \Re \left [ e^{-imt} \frac{\Theta (m)}{m} ( \sin md + \frac{\xi(m)}{m}( 1-\cos md)) e^{\xi (m) z} \right ]
\ee
and
\be
z\ge d, \ \ e(z,t)= J_0 \Re\left [ e^{-imt} \frac{\Theta (m)}{m} ( \sin md + \frac{\xi(m)}{m}( 1-\cos md)) e^{-\xi (m) (z-d)}\right ],
\ee
\end{widetext}
which is symmetric in $z\to d-z$. Notice that the electric field only penetrates inside the plates over a distance of order $1/\xi(m)$ corresponding to a skin effect. Finally,
we have for $ 0\le z\le d$
\begin{widetext}
\ba
 e(z,t)= J_0\Re \left [ e^{-imt} \frac{\Theta (m)}{m}  \left \{  \left (\sin m z+ \frac{\xi(m)}{m} (1-\cos m z) \right )
\left (\cos m(z-d) -\frac{\xi(m)}{m} \sin m(z-d) \right )
\right . \right .
\nonumber
\\
\left . \left .
+ \left (\cos mz + \frac{\xi (m)}{m} \sin mz\right )\left (\sin m (d-z) +\frac{\xi(m)}{m}\left ( 1- \cos m(d-z)\right )\right ) \right \}  \right ],
\nonumber \\
\ea
\end{widetext}
which is also symmetric in $z\to d-z$.

In the next section, we generalise this result to the case where the magnetic field is in all space, {\it i.e.} a more realistic situation, and calculate the classical Casimir pressure induced by the axion.

\subsection{Non-ideal conductors with magnetic field in all space}

\subsubsection{Electric field}

The configuration under consideration here is probably the closest to a possible realistic experimental setup. The cavity is made of conductors with finite conductivity and the magnetic field is present everywhere. The situation is sketched in Fig.~\ref{fig:twoplatesreal}.

\begin{figure}[h]
\centering
\includegraphics[width=\columnwidth]{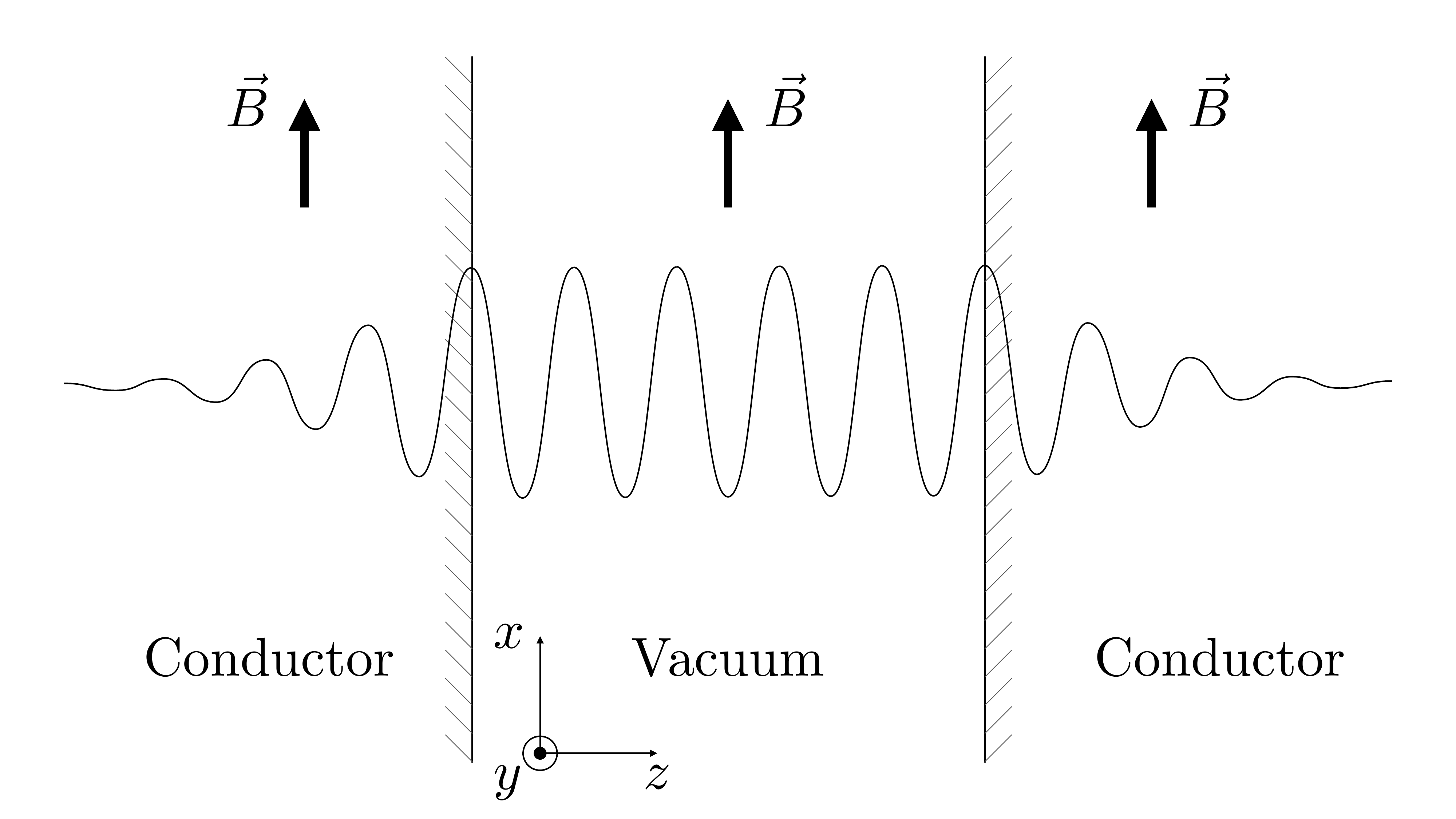}
\caption{Geometry and boundary conditions for the two-plate case with real conductors.}
\label{fig:twoplatesreal}
\end{figure}

When the magnetic field is present everywhere, the electric field receives new contributions $\delta e (z)$, which are given by
\begin{widetext}
\ba
\delta e(z,t)= J_0\Re \left [ e^{-imt+\xi(m)z}\frac{\Theta(m)}{\xi(m)}\left ( 1+e^{-\xi(m)z} \left (\cos md + \frac{\xi(m)}{m}\sin md \right )
\nonumber \right . \right .
\\
\left . \left .
+\frac{m}{\xi(m)}\left (1-e^{-\xi(m)z} \right ) \left (\sin md -\frac{\xi(m)}{m}\cos md \right) \right  )\right ] \nonumber \\
\ea
\end{widetext}
and the contribution obtained by $z\to d-z$ when $z\ge d$. Between the plates, for $z\in [0,d]$ the contribution is given by the following expression: 
\begin{widetext}
\ba
\delta e(z,t)=J_0 \Re \left [ e^{-imt} \frac{\Theta(m)}{\xi(m)}\left ( \cos m z + \cos m(d-z) + \frac{\xi(m)}{m}(\sin mz+ \sin m(d-z))\right )\right ] \nonumber , \\
\ea
\end{widetext}
which is symmetric in $z\to d-z$.

One interesting fact is that there is a non-vanishing oscillating electric field at infinity, which is not propagating and results from the coupling between the axion and the external magnetic field. More precisely we get far inside the plate
\begin{widetext}
\be
e_{\rm plate}(t)= J_0\Re \left [ e^{-imt}\frac{\Theta(m)}{\xi(m)}\left ( 2\cos md + \left (\frac{\xi(m)}{m}- \frac{m}{{\xi}(m)}\right )\sin md \right  ) \right ].
\ee
\end{widetext}
When the conductivity is larger than the mass, as in realistic situations, this simplifies to
\be
e_{\rm plaque}(t)\approx J_0\Re \left [ e^{-imt}\frac{\Theta(m)}{m }\sin md  \right ],
\ee
and finally
\be
e_{\rm plaque}(t)\approx -J_0\Re  \left [ \frac{e^{-imt}}{\xi^2(m)}  \right ] \simeq -J_0\Re \left [ i \frac{e^{-imt}}{m\sigma }  \right ].
\ee
This implies that there is a conduction current
\be
j_{\rm cond}(t)= \frac{J_0}{m}\sin mt,
\ee
the amplitude of which does not depend on the conductivity.

\subsubsection{Classical pressure }

The experimental observable for the effect under scrutiny here is a modification of the Casimir pressure in the void between the plates. The pressure is given by Eq.~\ref{eq:pressureaverage}, which comes only from the polarisation effects in matter given by
\be
p(d,t)= \frac{\vec B.\vec p}{B}= \int \slashed{d}\omega e^{-i\omega t} (\epsilon(\omega) -1) e(d,\omega).
\ee
As a result, we only need to evaluate the integral
\be
e(d,t)= J_0 \Re \left [ e^{-im t} \int_{-\infty}^\infty dz_0  G(m, d;z_0, 0) \right ].
\ee

The expressions that we need are given below.
When $z_0<0$ we have
\be
G(m,d;z_0)= \Theta(m) e^{\xi(m) z_0},
\ee
for $z_0\in [0,d]$,
\be
G(m,d;z_0)= \Theta (m) \left (\cos m z_0 + \frac{\xi(m)}{m}\sin m z_0\right ),
\ee
and finally
for $z_0>d$
\be
G(m,d;z_0)=\Theta (m) \left (\cos m d + \frac{\xi(m)}{m}\sin m d\right )e^{-\xi(z_0-d)},
\ee
which is continuous at $z_0=0,d$.
Now it is easy to see that  $\int_{-\infty}^\infty dz_0  G(m, d;z_0, t_0)$ converges and becomes
\begin{widetext}
\ba
 {\cal G}(m,d)=\int_{-\infty}^\infty dz_0  G(m, d;z_0, t_0)=
 \frac{\Theta(m)}{m}\left [\frac{m}{\xi(m)}\left (1+\cos md + \frac{\xi(m)}{m} \sin md \right )
 \right .
 \nonumber \\
 \left .
 + \sin md + \frac{\xi(m)}{m}(1-\cos md) \right ].
\ea
\end{widetext}
Then the electric field is given by
\be
e(\omega,d)= \frac{J_0}{2} \left ( {\cal G} (m,d)\slashed{\delta}(\omega-m) + \bar {\cal G} (m,d)\slashed{\delta}(\omega+m)\right ),
\ee
and the polarisability is
\be
p(t,d)= J_0 \Re \left [ e^{-imt} (\epsilon (m)-1) {\cal G}(m,d) \right ].
\ee
As a result  the pressure becomes
\be
\langle P_z\rangle = \frac{J_0^2}{2} \left \vert {\cal G}(m,d)\right \vert^2 \Re\left [\epsilon(m)-1\right ].\label{eq:axionpressure}
\ee
where
\be
\Re\left [\epsilon(m)-1\right ]= -\frac{\omega^2_{\rm Pl}}{m^2 +\gamma^2}.
\ee
Notice that this is always negative so the pressure is attractive. Moreover there is still a resonance when $\Theta$ diverges.

\section{Resonances in a Copper cavity}

For a fixed mass $m$ when the size of the cavity $d$ takes different values, the pressure on the plates experiences resonances. The resonances are due to the vanishing of $\Theta(m)$, which happens for complex values of $m$ with an imaginary part controlled by the parameter $\gamma$. In real situations, resonances occur for values close to $m_n= n\pi/d$  and their height is regularized by the dissipation parameter $\gamma$.
Effectively, the resonances on the real axis are obtained for
\be
\tan m_n d\simeq -\Re \left ( \frac{2}{\sqrt{-\epsilon(m_n)}}\right ),
\ee
where $\vert \epsilon(m_n)\vert \gg 1$. Close to the resonances, we have
\be
{\cal G}(m,d)\simeq \frac{2\Theta(m)}{m_n} \frac{\xi(m_n)}{m_n} \left (1-(-1)^n\right )
\ee
from which we recover that the resonances are only present when $n=2p+1$ \footnote{ Of course, the amplitude ${\cal G}$ does not vanish when $n=2p$. The subdominant terms are small compared to the resonant case when $n=2p+1$.}, this is a result we already obtained in the ideal case. For a fixed cavity size, the behavior of the pressure with the mass of the axion, is not trivial. This is illustrated in Fig.~\ref{fig:Pvsm}, for which the
 use of Copper plates is considered, corresponding to a damping time $\gamma$ parameter of
$10^{14}\rm \;Hz$, or
$6.6\times 10^{-11}\;\rm GeV$
 in natural units. We consider a 50~T magnetic field, close to the value of the higher stationary magnetic fields obtained experimentally, a plate separation of  $50\;\rm \mu m$ ($2.5\times 10^{11}\;\rm GeV^{-1}$ in natural units) and a generic axion scale of $10^{10}\;\rm GeV$.

\begin{figure}[h]
\centering
\includegraphics[width =\columnwidth]{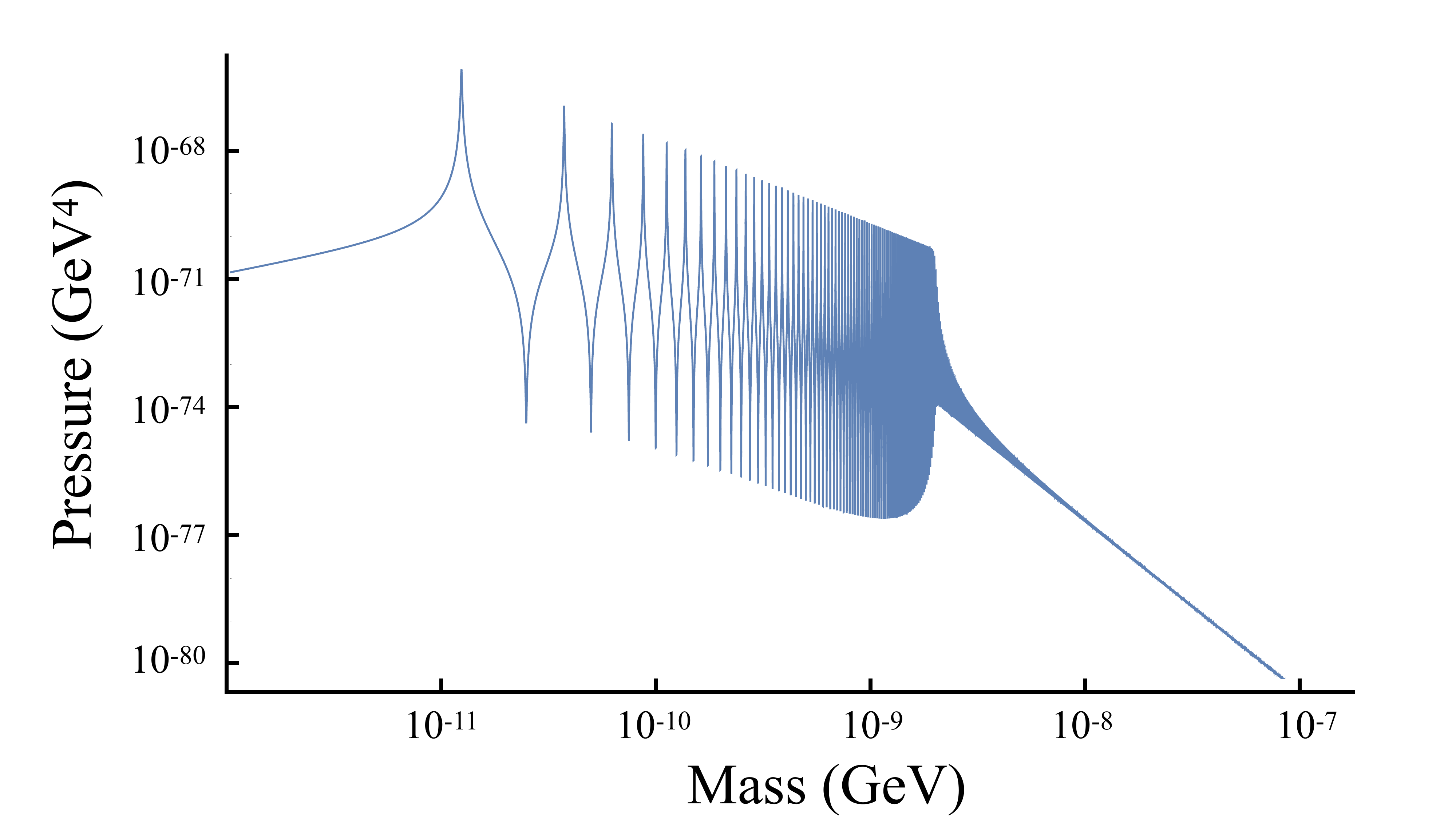}
\caption{The Casimir pressure (absolute value) from axions as a function of the axion mass for a distance $d=50\;\rm \mu m$ cm and $B=50$ T.}
\label{fig:Pvsm}
\end{figure}

In  Fig.~\ref{fig:Pvsm}, one can see the resonances, alternatively upward and downward. As expected the peak separation decreases as the mass increases. Resonances disappear gradually above $m=2\times10^{-9}\;\rm GeV$, which is precisely the plasma frequency for Copper.

For a fixed mass, the dependence of the axion classical pressure with the distance is displayed in Fig.~\ref{fig:PvsD}. The pressure varies on ~10 orders of magnitude between upward and downward peaks, around  a mean value that does not depend on the distance. In Fig.~\ref{fig:PvsD}, an axion mass of $1.26\times 10^{-11}\; \rm GeV$ is considered, corresponding to the first peak at $50\;\rm \mu m$.

\begin{figure}[h]
\centering
\includegraphics[width =\columnwidth]{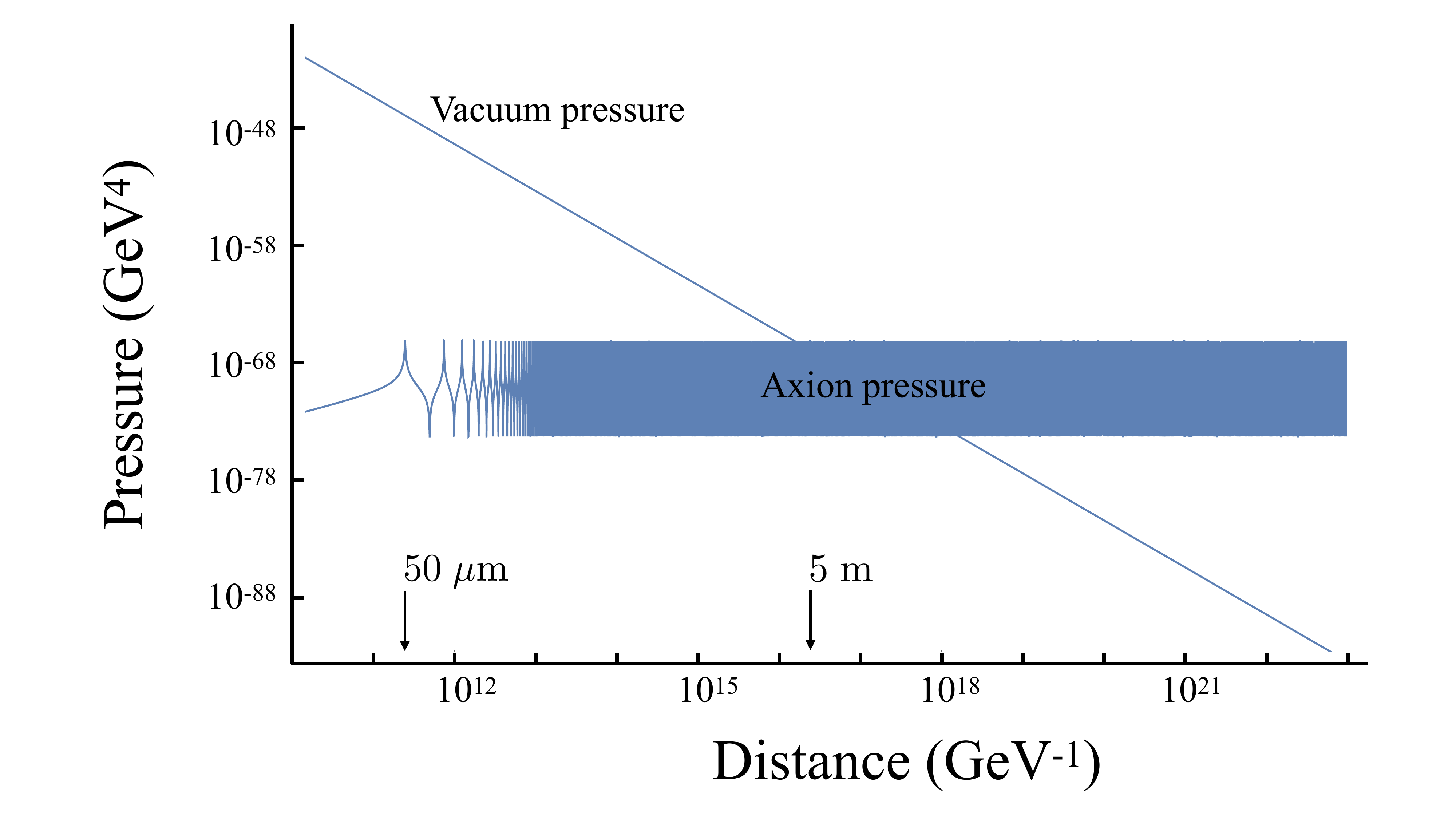}
\caption{The Casimir pressure (absolute value) from vacuum and from axions as a function of the distance between the metallic plates when a magnetic field $B=50$ T is present between the plates, and $m=1.26\times 10^{-11}\;\rm GeV$.}
\label{fig:PvsD}
\end{figure}

In Fig.~\ref{fig:PvsD}, the conventional vacuum pressure is shown in comparison to the axion pressure, with its typical $1/d^4$ dependance. At the first peak for $d=50\;\rm \mu m$, the vacuum pressure is 19 orders of magnitude higher than the axion pressure, making impossible to detect these axions in a Copper cavity. As the vacuum pressure drops very quickly with the size of the cavity, and as the axion field injects energy uniformly, the ratio becomes more favorable as the size increases. Around a few meters, some axion resonances are larger than the vacuum pressure. However, no measurement of Casimir pressure is available at this scale.

The search for axion through Casimir effects at short distances in a Copper cavity is not very promising as shown above. To search for a signal, the general rule is to try measuring the Casimir pressure on the largest possible distance. In addition, we will now see that the use of a reflecting material with a smaller value for the damping parameter $\gamma$ leads to more narrow resonances and might provide a higher signal.

\section{The damping parameter and the width of resonances}

In this section we investigate the influence of the damping parameter $\gamma$ on the strength of a potential axion signal.
From Eq.~\ref{eq:theta},  we can approximate
\be
\Theta(m) \simeq \frac{1}{2m_{2p+1}} \sqrt{-\epsilon(m_{2p+1})}\frac{1}{\frac{\sqrt{-\epsilon(m)}}{\Re\left (\sqrt{-\epsilon(m_{2p+1})}\right)} -1}
\ee
and upon using $\xi(m_n)/m_n \simeq \sqrt{-\epsilon(m_n)}$ we find that the Green's function reduces to
\be
{\cal G}(m,d)\simeq \frac{1}{m_{2p+1}^2}\frac{1}{\frac{\sqrt{-\epsilon(m)}}{\Re\left (\sqrt{-\epsilon(m_{2p+1})}\right )} -1}
\ee
close to the resonance at $m= m_{2p+1}$. This can be approximated in the two regimes corresponding to $m_{2p+1}\gg \gamma$ and $m_{2p+1}\ll \gamma$.

The width of the resonances will be controlled by the value of the damping parameters. Two regimes can be observed depending on the relative value of $\gamma$ and the axion mass. As the width diminishes, the peak signal naturally increases.

\subsection{Broad resonances $m\ll \gamma$}

We first consider when the resonances is given by $m_{2p+1}$ and when $m_{2p+1}\ll \gamma$.  In this case we have
\be
\sqrt{-\epsilon(m)} \simeq \frac{\omega_{\rm Pl}}{\sqrt {\gamma m}} e^{-i\pi/4}\left ( 1+\frac{i\omega}{2\gamma}\right ).
\ee
Notice that this is a complex number with a non-vanishing imaginary part even when $\gamma$ is vanishingly small. This implies that the width of the resonance has no direct
connection to $\gamma$ as can be seen in the approximation
\be
{\cal G}(m,d)\simeq \frac{1}{m_{2p+1}} \frac{1}{\frac{m-m_{2p+1}}{2} + i m_{2p+1}}.
\ee
As the width of the resonance is of order $m_{2p+1}$ and $m_{2p+1}\ll \gamma$, the width of the resonance is constrained to be
small compared to $\gamma$. On the other hand, as the width is of the order of the location $m=m_{2p+1}$ of the resonance, this corresponds to
a broad resonance. When several resonances of this type are present, the resonance of lowest order $p=0$ dominates as its amplitudes is the largest.
The resulting pressure becomes
\be
\langle P_z\rangle= - \frac{2\omega_{\rm Pl}^2 J_0^2}{\gamma^2 m^2_{2p+1} }\frac{1}{(m-m_{2p+1})^2 + 4m_{2p+1}^2}.
\ee
At the resonance we have
\be
\langle P_z\rangle= - \frac{2\omega_{\rm Pl}^2 J_0^2}{\gamma^2m^4_{2p+1}},
\ee
which shows that the strength of the signal is suppressed by $\gamma^2$ for high values of the damping parameter. So for large $\gamma$ values, the resonances are broader and with a reduced peak height. This is illustrated in Fig.~\ref{fig:broad}, where $\gamma = 10^{-8} \;\rm GeV$ has been used ($1.5\times 10^{16}$ Hz). It displays the first four resonances of the axion-induced pressure with respect to the distance between the plates.

\begin{figure}[h]
\centering
\includegraphics[width =\columnwidth]{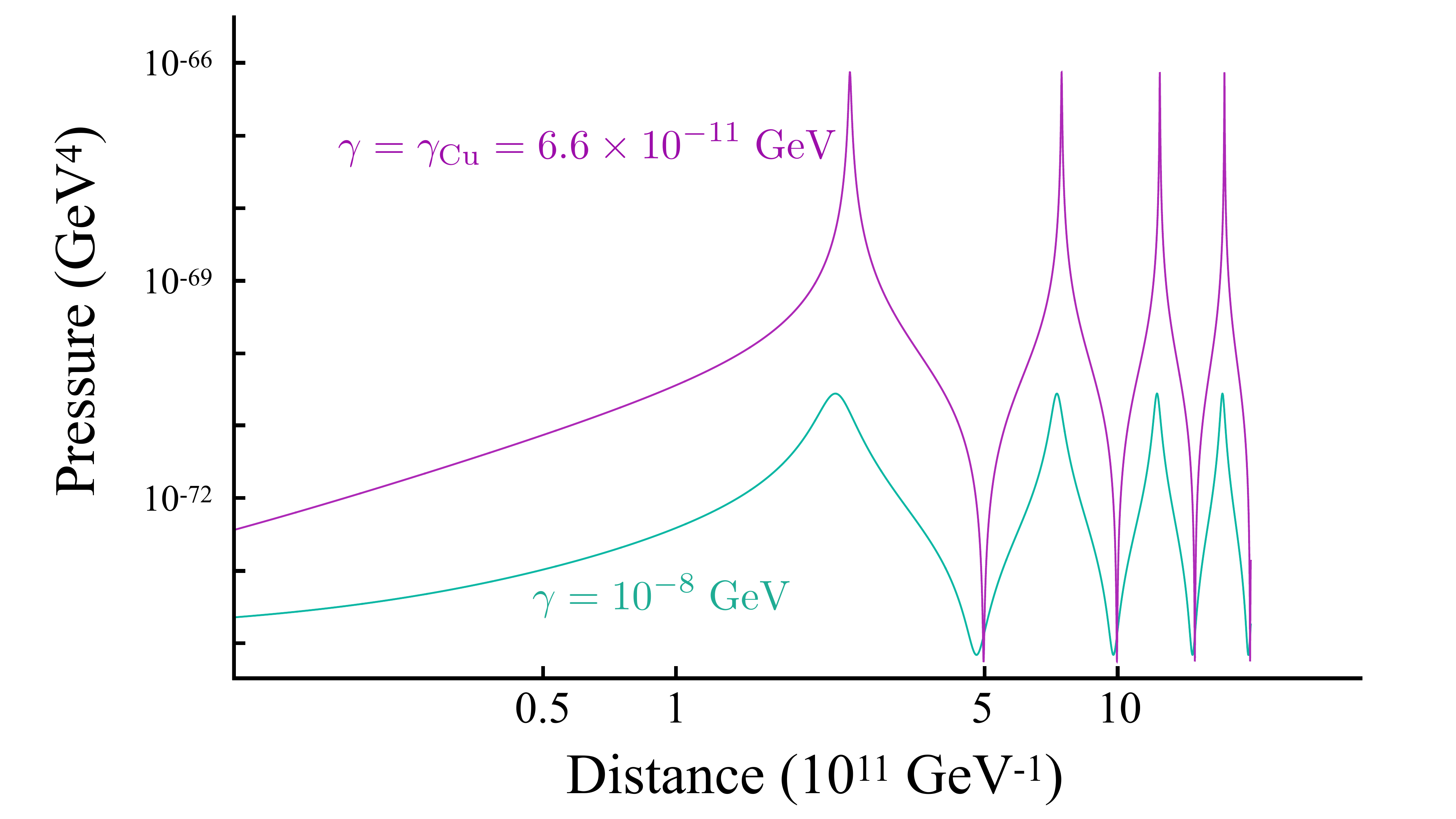}
\caption{First resonances from axions as a function of the distance between the metallic plates, with an increased value of the damping parameter $\gamma = 10^{-8} \;\rm GeV$, $B=50$ T, and $m=1.26\times 10^{-11}\;\rm GeV$.}
\label{fig:broad}
\end{figure}

\subsection{Narrow resonances $\gamma \ll m$}
It appears from the previous section that to increase the strength of the signal, it is better to consider a small damping parameter, {\it i.e.} a material that tends to be a better conductor.
We focus now on the case where $m_{2p+1}\gg \gamma$ for which we obtain
\be
\sqrt{-\epsilon(m)}\simeq \frac{\omega_{\rm Pl}}{\sqrt m}\left (1- \frac{i\gamma}{2m}\right ).
\ee
In this case, the imaginary part is directly related to $\gamma$.
This leads to
\be
{\cal G}(m,d)\simeq \frac{1}{m_{2p+1}} \frac{1}{ m -m_{2p+1} +\frac{i\gamma}{2}}.
\ee
This is a Breit-Wigner distribution with a width
$
\Gamma= \gamma
$,
much smaller than the location of the resonance at $m=m_{2p+1}$.
In this case we  obtain the pressure
\be
\langle P_z \rangle = - \frac{J_0^2}{2} \frac{\omega_{\rm Pl}^2}{m_{2p+1}^4} \frac{1}{(m-m_{2p+1})^2 + \frac{\gamma^2}{4}}.
\ee
At the resonance we have
\be
\langle P_z \rangle = - {2J_0^2} \frac{\omega_{\rm Pl}^2}{m_{2p+1}^4} \frac{1}{\gamma^2}.\label{eq:pressureresonn}
\ee
which diverges in the small $\gamma$ limit where the resonances become infinitely thin. Fig.~\ref{fig:narrow} shows how resonances get narrower and higher for small values of $\gamma$. For this figure, $\gamma = 10^{-17}\;\rm GeV$ (15 MHz) and all other parameters are unchanged compared to previous figures.

\begin{figure}[h]
\centering
\includegraphics[width =\columnwidth]{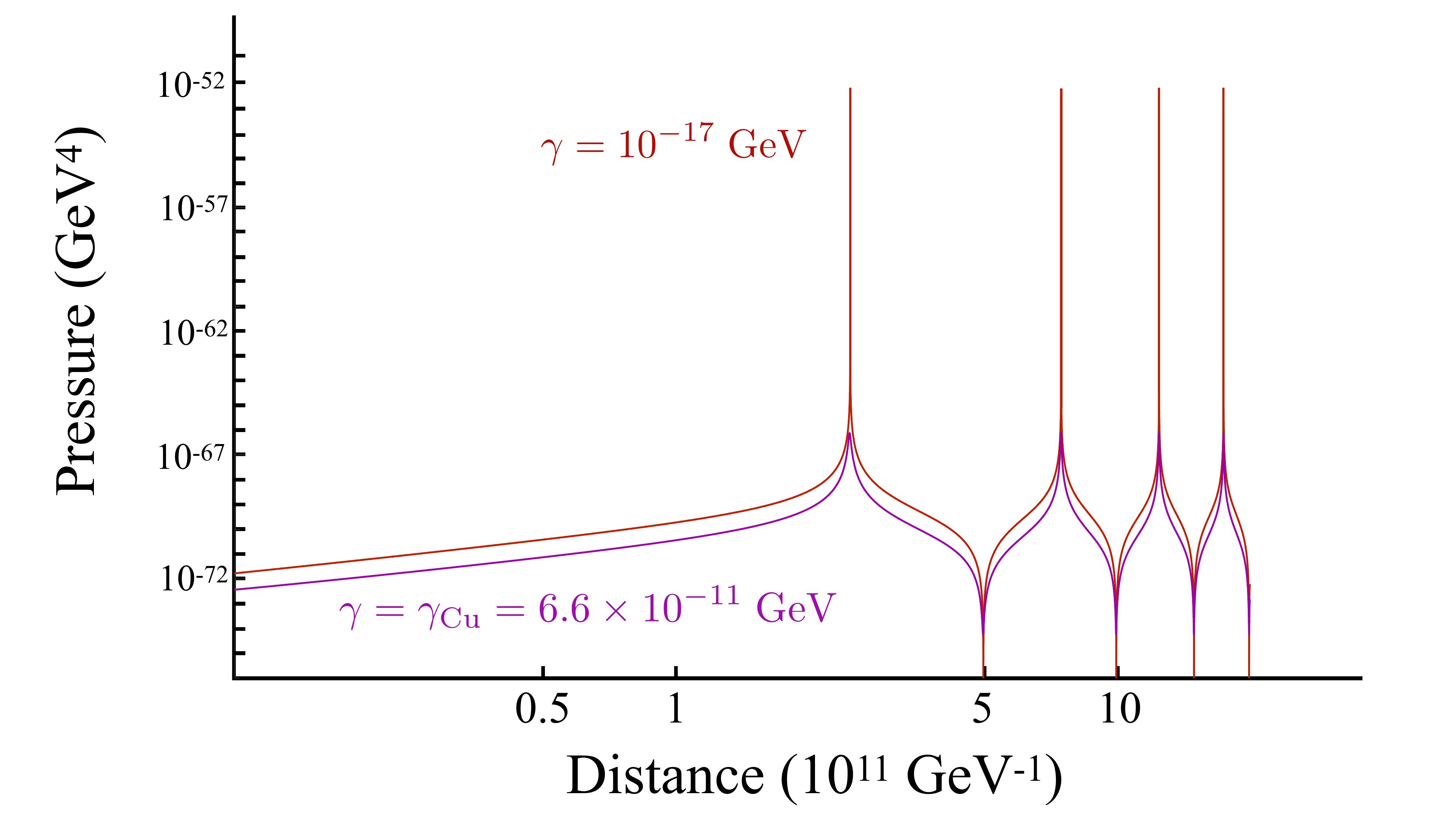}
\caption{First resonances from axions as a function of the distance between the metallic plates, with an decreased value of the damping parameter $\gamma = 10^{-17} \;\rm GeV$, $B=50$ T, and $m=1.26\times 10^{-11}\;\rm GeV$.}
\label{fig:narrow}
\end{figure}

\subsection{Comparison with the halo velocity distribution}

The width of the resonance is limited by the dispersion due to the finite velocity of dark matter in the galactic halo
\be
\frac{\Delta m}{m}= \frac{\gamma}{m} \gtrsim v^2.
\label{dispersion}
\ee
From an observational point of view, there is no interest in searching for a material with arbitrarily small $\gamma$. The natural velocity dispersion will blur the resonance anyway. As the Casimir effect cannot be seen for distances less than a typical distance $d_{\rm max}$ which is a fraction of a micron, we find that the observability of the axion effect compared to the usual Casimir effect will be possible only for
\be
\gamma \gtrsim \frac{\pi v^2}{d_{\rm max}}
\ee
For lower values of $\gamma$, the resonances would be affected by the velocity dispersion mostly. For $d_{\rm max}=50 \;\rm \mu m$, the limiting value of $\gamma$ is $\sim 10^{-17}\;\rm GeV$ (15 MHz).
Notice that the uncertainty on $m$ would also result in an uncertainty on the first resonance $\Delta d$ such that
\be
\frac{\Delta d}{d}= \frac{\gamma d}{\pi} \gtrsim v^2.
\ee



\section{Potential sensitivity of Casimir experiments}

To estimate the possible reach of Casimir experiments with a magnetic field, we assume an experiment could be built with a 50~T magnetic field  to perform a measurement of Casimir pressure between  $d=5\; \mu$m and  $d=50\; \mu$m.
A 50~T magnetic field is of the order of the highest intensity for stable magnetic fields produced in the laboratory, see for instance~\cite{nougat} (27~T) and SMHFF (45~T). Running a Casimir experiment in such a high intensity magnetic field is therefore probably challenging but not impossible.
We also assume that the plates are made out of an extremely good conductor and that the regularization of the resonances is the result of the natural velocity dispersion of the dark matter halo. This is independent of the material as long as the conductivity is high enough.
This could be achieved with superconducting material, or
any material with a damping factor of $\gamma \sim 10^{-17}\;\rm GeV$. The plasma frequency is taken equal to $2\times 10^{-9}\;\rm GeV$. As the London length for superconductors lie un the range 10 nm to 1000 nm, this seems achievable~\cite{techniquesdelingenieur}. We assume that a 1\% deviation of the vacuum Casimir pressure can be observed. To determine the sensitivity of such an experiment, we solve numerically the equation
\be
P_{\rm axion}\left (m,M\right)=\eta \times P_{\rm vacuum},
\label{eq:reach}
\ee
for the variables $m$ and $M$, where $\eta$ is the sensitivity of the experiment to the observation of a variation of the standard value of the pressure. In the following we assume $\eta=1\%$, meaning that a percent-level deviation to the conventional Casimir pressure is detectable. We consider the ideal pressure relation
\be
P_{\rm vacuum} = - \f{\pi^2}{240 d^4},
\ee
so by taking the value of the axion-induced pressure at a resonance~\ref{eq:pressureresonn} for a distance $d$ one gets the value of the sensitivity on the axion-photon coupling
\be
g_{\rm sens} = \sqrt{\eta}\f{\pi}{\sqrt{960}}\f{\gamma}{\omega_{\rm P}}\f{m}{B d^2} ,
\ee
where the coupling is simply taken as the inverse of the axion scale $g=1/M$.

The projected constraints are displayed in Fig.~\ref{fig:reach} in a coupling-mass plane. The mass range is imposed by the range of distances over which the Casimir effect can reasonably be probed. There exist several constraints in the considered mass range, at low mass, the laboratory experiments ALPS~\cite{Ehret_2010} and PVLAS~\cite{Ejlli:2020yhk} constraint the highest values of the coupling. In the whole range, the blue region is the constraint from the CAST helioscope~\cite{cast2017}. The blue regions on the right side of the plot are direct constraints from telescopes (JWST~\cite{janish2023hunting, roy2023sensitivity}, WINERED~\cite{yin2024result}, MUSE~\cite{Regis_2021}, VIMOS~\cite{Grin_2007}), and indirect constraints from the level of infrared background in the universe as measured by $\gamma$-rays~\cite{Bernal_2023}.
The Casimir type of experiments for axion search obviously require that axions are the dark matter. Above the dashed red line labelled ``"decay limit", in the redish region the lifetime of the axions is shorter than the age of the universe so they cannot be the dark matter. Finally The oblique yellow band labelled ``QCD'' corresponds to the region where parameters nicely combine to solve the strong-CP problem.
The main result of this paper is the central light-green region labelled ``Casimir $5\;\rm\mu m - 50 \;\mu m$''. It
contains the parameters that comply with the condition~\ref{eq:reach} for $\eta=1\%$. The edge at 2 eV is due to the value of the plasma frequency, above which resonances disappear.

To understand the shape of the sensitivity curve that is shown below, let us consider the first resonance. At short distance, the ratio between the vacuum pressure and the axion pressure is not favorable, because $P_{\rm vacuum}$ decreases very steeply with $d$ and $P_{\rm axion}$ is constant on average. The mass corresponding to this first resonance is as high as it can get because of the inverse proportionality between $m_{\rm a}$ and $d$ at resonance. One can imagine an experiment where $d$ can be varied. When $d$ increases, two things happen : the mass of the first resonance decreases, and the ratio $P_{\rm axion}/P_{\rm vacuum}$ improves, making the sensitivity better. In a coupling/mass plane, going from short distance to large distance, the sensitivity follows a line from a larger mass / larger coupling value, to a lower mass lower coupling value. The same reasoning is true for all the other resonances, thus leading to a sawtooth-like line for the sensitivity curve of Casimir experiments. This is what appears in Fig.~\ref{fig:reach}. The figure also displays various existing constraints in the axion coupling/mass plane, the excluded regions being the different blue patches.
\begin{figure}[h]
\centering
\includegraphics[width =\columnwidth]{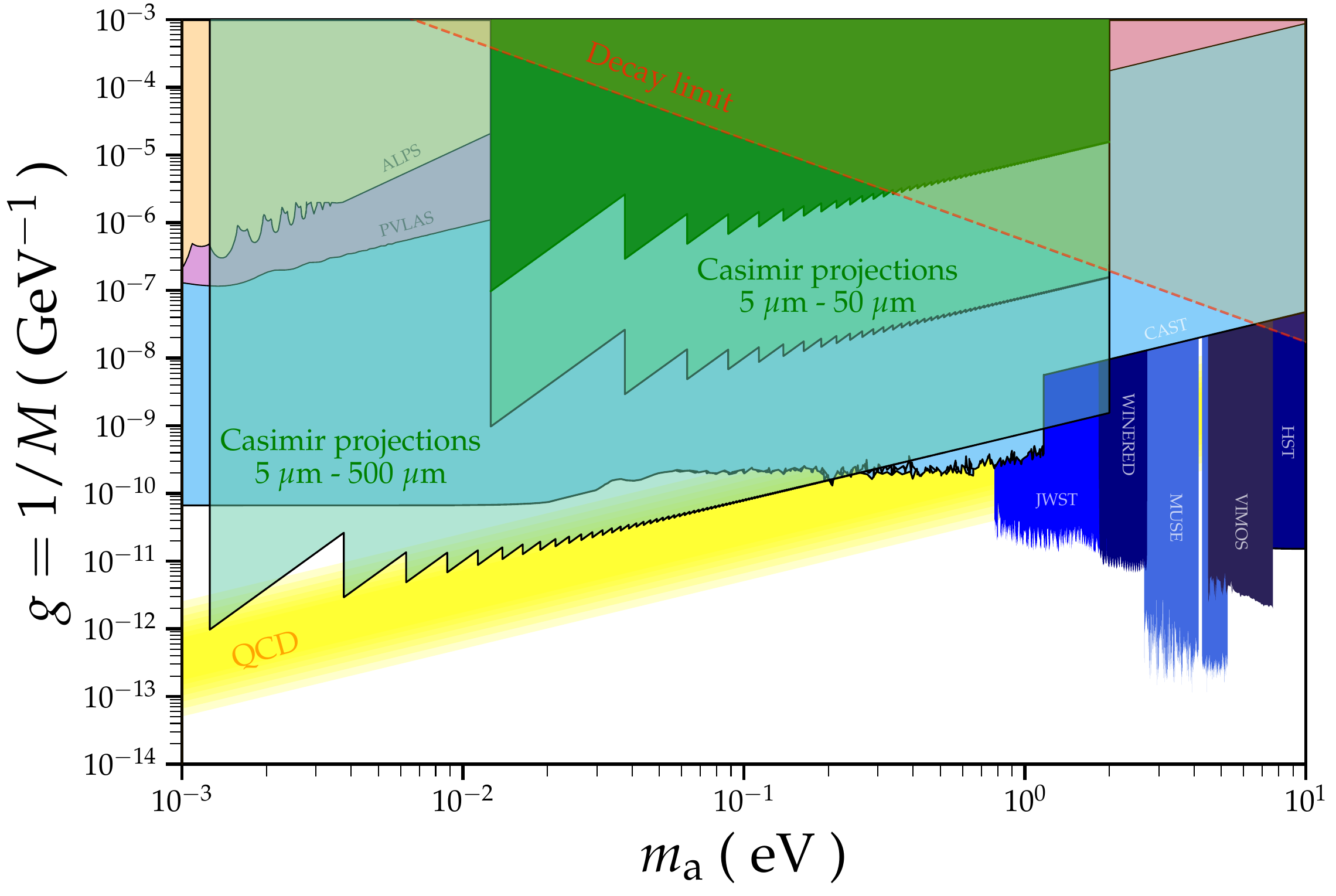}
\caption{Potential reach of Casimir experiments in a mass / coupling plane. The three green regions correspond to different assumptions on the experimental parameters (see text).}
\label{fig:reach}
\end{figure}

To show the limits of the current sensitivity estimates, some parameters can be varied. For example it is expected that the plasma frequency in superconducting materials depend on the magnetic field~\cite{quantum3040046}. In that case the plasma frequency could be lower.
In Fig.~\ref{fig:reach}, the dark green region at higher values of the coupling corresponds to resonances of the same width as for the other curves, but with higher values of the plasma frequency, here $\omega_{\rm p} = 2\times 10^{-11}\;\rm GeV$.
On a more optimistic side, If it were possible to measure the Casimir force on distances up to 500 $\mu$m, the sensitivity region would enlarge towards lower masses and lower coupling values, as shown in lighter green in Fig.~\ref{fig:reach}. If it were possible to probe the Casimir pressure on such large distances, some parameters favored by QCD could be tested.

 Finally let comment briefly on the quantum contribution to the Casimir pressure from the presence of the rapidly  oscillating axion field. The correction to the Casimir pressure can be estimated by consider a typical one loop vacuum diagram for the photons with two axionic insertions.  At each vertex and working in the Coulomb gauge, one time derivative acts on the oscillating axion leading to a factor of $m^2$. The diagram itself is proportional to $\phi_0^2/M^2$. Finally a factor of $d^2$ must be inserted for dimensional reason, see the appendix \ref{app:quantum} for a sketch of the calculation. The fact that the contribution vanishes when $m=0$ follows from the fact that for a constant axion field, the $F\tilde F$ term in the Lagrangian is non-dynamical and therefore no pressure can be generated. All in all we have
\be
\frac{\delta P_{\rm axion}}{P_{\rm casimir}} = {\cal O}\left (\frac{\phi_0^2}{M^2} m^2 d^2\right )
\ee
which is negligible for the typical values considered in this paper.  A more thorough analysis is left for the future.

\section{Conclusion}

We showed that the presence of axionic dark matter in the form of an oscillating scalar field coupled to electromagnetism modifies the Casimir effect at the classical level, i.e. it induces a classical pressure between metallic plates. The main new phenomenon  resulting from our study is the appearance of a series of resonances whose  positions depend on the spatial scale probed by the Casimir experiment. The resonances are regularized by the finite conductivity of the metallic plates. Estimate of the sensitivity show that the potential reach of this method could be competitive with other probes such as helioscopes or haloscopes. This will require experiments sensitive to long distance Casimir forces, and very high magnetic fields. The comparison between the classical pressure calculated here and the Casimir pressure expected between the two plaques will have to take into account the finite temperature effects and also most importantly the newly discovered strong effects that large background magnetic fields have on the quantum Casimir pressure \cite{Zhang}. This is left for future work.

\acknowledgements{
This project has received funding from the European Research Council (ERC) under the
European Union's Horizon 2020 research and innovation programme (Grant agreement No.~865306). We would like to thank Andreas Ringwald for relevant remarks and suggestions.
}

\bibliography{BraxBrun_Casimir}

\appendix

\section{Radiation from a single plate}
\label{ap:single}

As an application of the Green's function techniques, we will derive the expression for the electric field when only one plate is present and the electric field penetrates inside a metal with finite conductivity. This generalises the ideal case where the conductivity is taken to be infinite.
We first consider the generation of an electric due to the axion coupling when the magnetic field occupies all space. This is given by
\be
e(z,t)= J_0 \Re \left [ e^{-im t} \int_{-\infty}^\infty dz_0  G(m, z;z_0, t_0) \right ].
\ee
where the Green's function is such that it vanishes in vacuum at $z=+\infty$ and in matter at $z=-\infty$.
In vacuum, this selects the modes in $e^{i\omega z}$ where the replacement $\omega^2 \to \omega^2 +i\tilde \epsilon$ with $\tilde \epsilon \to 0$ guarantees the convergence.

We now have to select the Green's function for $z>0$, i.e. outside the plaque. This gives
\be
z_0<0: \ G(\omega,z;z_0)= \frac{e^{i\omega z +\xi(\omega) z_0}}{\xi(\omega) -i\omega},
\ee
and
\begin{widetext}
\be
z_0\in [0,z]: G(\omega,z;z_0)= \left ( \frac{1}{i\omega +\xi(\omega)}-\frac{1}{2i\omega}\right )e^{i\omega (z+z_0)}+ \frac{e^{i\omega (z-z_0)}}{2i\omega},
\ee
\end{widetext}

whilst finally
\begin{widetext}
\be
z_0>z: G(\omega,z;z_0)=\left ( \frac{1}{i\omega +\xi(\omega)}-\frac{1}{2i\omega}\right )e^{i\omega (z+z_0)}+ \frac{e^{i\omega (z_0-z)}}{2i\omega}.
\ee
\end{widetext}

As a result we have
\begin{widetext}

\be
\int_{-\infty}^\infty dz_0  G(m, z;z_0, 0)= \frac{1}{m^2} +\left ( \frac{1}{\xi(m)(\xi(m)-im)}- \frac{\xi(m)}{m^2}\frac{1}{\xi(m)+im}\right )e^{imz}.
\ee
\end{widetext}

Hence there are two components. There is an oscillating electric field at the frequency $m$ of the axion oscillations. There is also a propagating wave with frequency $m$
\begin{widetext}
\be
e(z,t)= \frac{J_0 \cos mt}{m^2}+ J_0 \Re \left [ \frac{1}{\xi(m)(\xi(m)-im)}- \frac{\xi(m)}{m^2}\frac{1}{\xi(m)+im} e^{im(z-t)}\right ].
\ee
\end{widetext}

Let us now assume that the magnetic field is only in the plaque for $z<0$, then the external electric is obtained by evaluating
\be
\int_{-\infty}^0 dz_0  G(m, z;z_0, 0)=  \frac{1}{\xi(m)(\xi(m)-im)}e^{imz},
\ee
corresponding to a propagating wave
\be
e(z,t)= J_0 \Re\left [\frac{1}{\xi(m)(\xi(m)-im)}e^{im(z-t)}\right ].
\ee
As typically $\xi(m) \sim \omega_{\rm Pl}$ for $m\ll \omega_{\rm Pl}$, we have the order of magnitude estimate
\be
e(z,t)\simeq \frac{J_0}{\omega_{\rm Pl}^2} \cos (m(z-t),
\ee
i.e. a propagating wave of small amplitude.

In the case the magnetic field does not penetrate in the plaque we have
\be
e(z,t)= \frac{J_0 \cos mt}{m^2}- J_0 \Re \left[ \frac{\xi(m)}{m^2}\frac{1}{\xi(m)+im}e^{im(z-t)}\right ],
\ee
where we see that the field vanishes at the surface only in the ideal case.

\section{The  Green's function in the two-plate case}
\label{app:G}

In this appendix we will give explicit formulae for the Green's function satisfying
\be
-\partial_0( \epsilon\star_t \partial_0 G) + \Delta G=  \delta^{(4)}(x^\mu-y^\mu).
\ee
As explained in the main text, as we are considering classical solutions to a field configuration depending on $z$ along the external field in the $x$ direction, we only need the scalar Green's function for the Maxwell equation on the $x$ direction. The full tensorial nature of the problem is discussed in chapter 81 of \cite{Landau9} where the absence of off-diagonal terms can be ascertained by taking $q=0$ in eq. (81.5) for instance. For the sake of generality, we give here the Green's function for the scalar Maxwell equation where we restore the dependence on $(x,y)$. In the main text we only need the following expressions with $p_\parallel=0$. The reader only interested in applications can take these expressions for granted or refer to \cite{Landau9} for details.
In the following we will use time-translation invariance and space-translation invariance in the $(x,y)$ plane along the plaques  to choose $y^\mu=(\vec 0,z_0,0)$.
In terms of Fourier decomposition the Green's function satisfies
\be
\left (\partial_z^2 -p_\parallel^2 + \epsilon(\omega) \omega^2\right ) G= \delta(z-z_0).
\ee
We define
\be
\xi= \left ( p_\parallel^2 -\epsilon(\omega)\omega^2\right )^{1/2},
\ee
where the square root is such that the real part is always positive. We also have
\be
\Delta= \left (\omega^2-p_\parallel^2 \right )^{1/2}.
\ee
We will separate the $z$-axis into three intervals and give the solution in each case using the continuity of the Green's function at the boundaries
$z=0$ and $z=d$. The first derivative is continuous there too whilst there is a jump $[\frac{dG}{dz}\vert_{z=z_0}=1$ of the first derivative when $G$ itself is continuous at $z=z_0$. Moreover we impose that $\lim_{\vert z\vert \to \infty} G=0$. This determines a unique solution.

\subsection{$z_0\in [0,d]$}
When $0\le z_0 \le d$ is between the plates
the Green's function is then defined by
\be
G=G_- e^{\xi z},\ z<0
\ee
and
\be
G= G_+ e^{-\xi(z-d)}, \ z>d,
\ee
where we have now
\be
G_-= \Theta(\omega,p_\parallel)  \left (( \cos \Delta  (z_0-d) - \frac{\xi}{\Delta} \sin \Delta (z_0-d)\right )
\ee
and
\be
G_+= {\Theta (\omega,p_\parallel)} \left ( \cos \Delta z_0 + \frac{\xi}{\Delta} \sin \Delta z_0\right )
\ee
and the Green's function between the plates
\begin{widetext}
\ba
&& 0\le z\le z_0, \ G= G_- \left ( \cos \Delta z +\frac{\xi}{\Delta} \sin \Delta  z \right )\nonumber \\
&& z_0 \le z \le d, \ G= G_+ \left ( \cos \Delta (z-d) -\frac{\xi}{\Delta} \sin \Delta (z-d)\right ),\nonumber \\.
\ea
\end{widetext}
where
\be
\Theta (\omega,p_\parallel)= \frac{1}{\Delta \left (\left  (1- \frac{\xi^2}{\Delta^2}\right ) \sin \Delta  d -2 \frac{\xi}{\Delta} \cos \Delta d\right )}.
\ee

\subsection{$z_0<0$}
This case is different from the previous case as
\be
G=G_- e^{\xi z},\ z\le z_0,
\ee
and
\be
G= A e^{\xi z} + B e^{-\xi z}, \ z\in [z_0,0],
\ee
whilst
\begin{widetext}
\ba
&& 0\le z\le d, \ G= G_+ \left ( \left (\sin \Delta d -\frac{\xi}{\Delta} \cos \Delta d \right ) \sin \Delta z+ \left (\cos \Delta d+ \frac{\xi}{\Delta} \sin \Delta d\right ) \cos \Delta z \right )\nonumber \\
&&  z\ge d, \ G= G_+ e^{-\xi (z-d)} \nonumber \\.
\ea
We find that
\ba
&&A= \frac{G_+}{2} \left ( \cos \Delta d+ \frac{\xi}{\Delta} \sin \Delta d + \frac{\Delta}{\xi} \left (\sin \Delta d -\frac{\xi}{\Delta} \cos \Delta d\right )\right )\nonumber \\
&&B=\frac{G_+}{2}\left ( \cos \Delta d+ \frac{\xi}{\Delta} \sin \Delta d - \frac{\Delta}{\xi}\left (\sin \Delta d -\frac{\xi}{\Delta} \cos \Delta d\right )\right )\nonumber\\
\ea
\end{widetext}
and we have
\be
G_+= \Theta(\omega,p_\parallel) e^{\xi z_0}.
\ee
Finally we find that
\begin{widetext}
\be
G_-=\Theta(\omega,p_\parallel)\left (\cosh \xi z_0\left  (\cos \Delta d+ \frac{\xi}{\Delta} \sin \Delta d \right )+ \frac{\Delta}{ \xi} \sinh \xi z_0 \left (\sin \Delta d -\frac{\xi}{\Delta} \cos \Delta d\right )\right )
\ee
\end{widetext}
\subsection{$z_0>d$}
We have now
\be
z<0,\ G= G_- e^{\xi z}
\ee
and
\be
z \in [0,d],\ G= G_-\left ( \cos \Delta z +\frac{\xi}{\Delta} \sin \Delta z\right),
\ee
whilst we have
\begin{widetext}
\be
z\in [d,z_0], \ G= \frac{G_-}{2} \left ( \sin \Delta d \left( \frac{\xi}{\Delta} + \frac{\Delta}{\xi}\right )e^{-\xi (z-d)} +\left [2 \cos \Delta d+ \sin \Delta d \left (\frac{\xi}{\Delta}-\frac{\Delta}{\xi}\right )\right ]e^{\xi(z-d)}\right )
\ee
\end{widetext}
and finally
\be
z>z_0, \ G= G_+ e^{-\xi( z-d)},
\ee
where we find
\be
G_-= \Theta (\omega, p_\parallel) e^{-\xi(z_0-d)}
\ee
and
\begin{widetext}
\be
G_+=\Theta (\omega, p_\parallel)\left (\cosh \xi (z_0-d) \left (\cos \Delta d+ \frac{\xi}{\Delta} \sin \Delta d \right )+ \frac{\Delta}{ \xi} \sinh \xi (z_0-d)\left (-\sin \Delta d +\frac{\xi}{\Delta} \cos \Delta d\right )\right ).
\ee
\end{widetext}

\section{Dissipation}
\label{sec:app_dissipation}

In the background magnetic field,  the complete equations of motion for $a^\mu$ can be deduced from the complete action \footnote{Notice that this action is only gauge invariant $a_\mu \to a_\mu + \partial_\mu \alpha $ on shell when we impose current conservation
$\partial_\mu {\cal J}^\mu=0$. This can be remedied by introducing a St\"uckelberg field $\theta$ as
\be
S= \int d^4 x \left (-\frac{1}{4} f_{\mu\nu}f^{\mu\nu}- \frac{\phi}{4M} f^{\munu}\tilde f_{\mu\nu}  - {\cal J^\mu}( a_\mu -\partial_\mu \theta)\right )
\ee
which transforms as $\theta \to \theta +\alpha$ under a gauge transformation. The equations of motion of $\theta$ give $\partial_\mu {\cal J}^\mu=0$ whilst the value of $\theta$ is left undetermined. This does not matter as the Hamiltonian obtained by Legendre transform with respect to both $a^\mu$ and $\theta$ does not depend on $\theta$.}
\be
S= \int d^4 x \left (-\frac{1}{4} f_{\mu\nu}f^{\mu\nu}- \frac{\phi}{4M} f^{\munu}\tilde f_{\mu\nu}  - {\cal J}^\mu a_\mu\right ).
\ee
The magnetic field  induces the  source term for the gauge field. It is convenient to rewrite the action in terms of the electric and magnetic fields using
$
-\frac{1}{4} f_{\mu\nu}f^{\mu\nu}= \frac{1}{2}\left (\vec e^2 -\vec b^2\right )
$
and
$
-\frac{1}{4}f^{\munu}\tilde f_{\mu\nu}=  \vec e.\vec b
$
resulting in the action
\be
S= \frac{1}{2}\int d^4 x \left (\vec e^2 -\vec b^2+ \frac{2\phi}{M} \vec e. \vec b - 2\vec {J}.\vec  a\right ),
\ee
from which the equations of motion used in the main text can be obtained.

Inside matter as in the case of metals, the physics is modified.  The permittivity $\epsilon$ of the material and its effect on the electric field captures in an effective way the interactions between the matter particles and the photons. This leads to the existence of the  displacement vector which is related to the polarisation of the medium by $\vec d= \vec e + \vec p$. The displacement vector is related to the electric field via a retarded effect
\be
\vec d(\vec x, t) = \int d\tau \epsilon(t-\tau) \vec e(\vec x, \tau).
\ee
A naive action due to the presence of a non-trivial permittivity can be postulated as
\be
S= \frac{1}{2}\int d^4 x \left (\vec e.\vec d -\vec b^2+ \frac{2\phi}{M} \vec e. \vec b - 2\vec {J}.\vec a\right ).
\label{toto}
\ee
As the action is not local anymore, the concept of particle is ill-defined. Moreover dissipation coming from the imaginary part of $\epsilon (\omega)$ breaks unitarity.
Let us use this action naively and get  the canonical momentum associated to the vector potential
\be
\vec \pi= -\frac{1}{2}(\epsilon+ \hat \epsilon)\star_t \vec e  - \frac{\phi}{M} \vec b,
\ee
where we have introduced the notation $\hat \epsilon (t)= \epsilon (-t)$ and $\star_t$ is the convolution operator in time.
The Hamiltonian becomes
\be
H= \frac{1}{2}\int d^3 x  \left (\vec e.\vec {\hat d} +\vec b^2+ 2\vec { J}. \vec a\right ).
\ee
where we have defined
\be
\vec {\hat d}= \hat \epsilon\star_t \vec e= \int d\tau \epsilon (\tau-t) \vec e(\tau)
\ee
As can be seen this differs from the usual energy postulated for electromagnetism
\be
H_{\rm mat}= \frac{1}{2}\int d^3 x  \left (\vec e.\vec { d} +\vec b^2+ 2\vec { J}. \vec a\right ).
\ee
The two Hamiltonians reduce to the same expression when $\epsilon=\hat \epsilon$. This condition can be understood
after Fourier transforming in time
$
\vec d(\omega, \vec x)= \epsilon (\omega) \vec e(\omega, \vec x)
$
where $\epsilon(\omega)$ is a complex function such that $\bar \epsilon (-\omega)= \epsilon (\omega)$. In a metal, the permittivity has a pole on the negative  imaginary axis.
There is dissipation unless $\epsilon (-\omega)= \epsilon (\omega)$ implying that $\Im \epsilon=0$, i.e. the imaginary part of $\epsilon (\omega)$ vanishes. When this is the case we have $\epsilon(t)=\epsilon (-t)$ and the Hamiltonian $H$ coincide with $H_{\rm mat}$. In general this is not the case and the two Hamiltonians differ.

There is another major issue with the naive action: it does not reproduce the correct equations of motion. Indeed the Euler-Lagrange for $\vec a$ coming from (\ref{toto}) is
\be
-\partial_0 \left ( \frac{(\epsilon+\hat \epsilon)}{2} \star_t \partial_0 \vec a\right ) + \Delta \vec a=- \frac{\dot \phi}{M} \vec b + \vec { J}
\label{wrong}
\ee
which does not coincide with the phenomenological equation of motion
\be
-\partial_0 ( \epsilon \star_t \partial_0 \vec a) + \Delta \vec a=- \frac{\dot \phi}{M} \vec b + \vec { J}.
\label{dis1}
\ee
This equation is not time-reversal invariant as $\epsilon\ne \hat \epsilon$, i.e. dissipation implies a specific arrow of time. In fact it is interesting to consider the time-reversed process defined as
\be
\tilde a^i(\tilde t)= a^i(-t),
\ee
where $\tilde t= -t$. It satisfies the time reversed equation
\be
-\partial_0 ( \hat \epsilon \star_{\tilde t} \partial_0 \vec {\tilde a}) + \Delta \vec {\tilde a}=- \frac{\dot \phi}{M} \vec b + \vec { J}.
\label{dis2}
\ee
where time derivatives are with respect to $\tilde t$.
The equation of motion (\ref{wrong}) is the symmetrised in time of (\ref{dis1}) and (\ref{dis2}). In this case when $\epsilon(t)= \epsilon(-t)$ the normal process and its time-reversed satisfy the same equation.  This only applies to plasmas and not to metals with a finite conductivity.

\section{Energy budget}\label{ap:budget}

\subsection{The small mass limit in the one plaque case}
Let us consider the case of $m\ll \gamma\ll \omega_{\rm Pl}$ then we have
\be
\xi \simeq \sqrt{m\sigma} e^{-i\pi/4}\left ( 1+ i\frac{m}{2\gamma}\right ).
\ee
This implies that
\begin{widetext}
\be
\frac{1}{m^2} +  \frac{1}{\xi(m)\left (\xi(m)-im\right )}- \frac{\xi(m)}{m^2}\frac{1}{\xi(m)+im}\simeq \frac{-i}{m\sqrt{m\sigma}}e^{i\pi/4}.
\ee
When the magnetic field is present everywhere
\be
e(z,t)= \frac{J_0}{m^2} \cos mt -\frac{J_0}{m^2}\cos m(t-z) + \frac{J_0}{m\sqrt{m\sigma}}\sin\left ( m(t-z) -\frac{\pi}{4}\right),
\ee
\end{widetext}
corresponding to the electric field generated in the ideal case complemented by a term which disappears when the conductivity because infinite.
The pressure on the plaques follows from
\be
\Re \left [\epsilon(m)-1 \right ]= - \frac{\omega^2_{\rm Pl}}{\gamma^2 +m^2}
\ee
and become
\be
\langle P\rangle \simeq- \frac{J_0^2}{2\gamma m^3 },
\ee
which is always negative, corresponding to the fact that the vacuum for $z>0$ attracts the plate situated at $z\le 0$.

In the single plaque case with a magnetic field only in the metal we have
\be
e(z,t)= \frac{J_0}{m\sigma }\sin m(t-z),
\ee
which vanishes for a perfect conductor.
The pressure on the plaque becomes
\be
\langle P\rangle\simeq  \frac{J_0^2}{2m^3 \gamma },
\ee
i.e. the same result as before when the magnetic field is everywhere.

In both cases the vacuum attracts the plate. This effect will reemerge in the two plate case where the two plates will attract each other classically.

\subsection{Energy balance}

It is interesting to consider the conservation of energy when one plaque is present and radiation from the plate takes place.
Inside the plaques we have the Maxwell equation
\be
\vec \nabla \wedge \vec b = \partial_0 \vec e + \vec j_{\rm ind}+ \vec j_{\rm axion},
\ee
where $\vec j_{\rm ind}$ is the induced current
\be
\vec j_{\rm ind}= \dot {\vec p},
\ee
where $\vec p$ is the polarisability.
This current only exists in a finite width of the plaque, i.e. this is a skin effect.
This current is responsible for the dissipation power $\vec j_{\rm ind}. \vec e$ in the plaque corresponding to the loss of energy. As a result, the plaque  heats up due to the Joule effect.
The axionic current is given by
\be
\vec j_{\rm axion}= \frac{\dot \phi}{M} \vec B.
\ee
This is complemented with the Bianchi identity
\be
\partial_0 \vec b +\vec \nabla \wedge \vec e=0,
\ee
from which we can get the propagation equation in matter for the magnetic field
\be
\left (\epsilon(\omega) \omega^2 + \Delta\right )\vec b=0.
\ee
It coincides with the propagation equation for the electric field.

Let us now study the local conservation of energy
\be
E= \frac{\vec e^2 +\vec b^2}{2},
\ee
which satisfies
\be
\frac{d E}{dt}+\vec \nabla . \vec P_{\rm poynting}=- \left (\vec j_{\rm ind}+\vec j_{\rm axion}\right ). \vec e,
\ee
where the Poynting vector corresponds to the radiated power
\be
\vec P_{\rm poynting}= \vec e\wedge \vec b.
\ee
In the absence of currents we have
\be
\frac{d E}{dt}+\vec \nabla . \vec P_{\rm poynting}=0,
\ee
expressing the conservation of energy, i.e. the energy lost by the system is matched by the flux of radiation. When the axion is present and dissipation too, we find that the power dissipated by Joule's effect is
\be
P_{\rm joule}= \langle \vec j_{\rm ind}.\vec e \rangle,
\ee
which has to be positive. We will study this local equation more precisely below.

In the case treated in the main text with a single plaque and a magnetic field
the polarisability at the surface of the metal does not vanish and becomes
\be
p(0,t)=  -\frac{J_0\sigma \cos \left (mt-\frac{\pi}{4}\right )}{m^2\sqrt{m\sigma}}
\ee
in the small mass limit.
This induces a surface current
\be
j_{\rm ind}(0,t)= \frac{J_0\sigma \sin \left (mt-\frac{\pi}{4}\right )}{m\sqrt{m\sigma}}.
\ee
The power dissipated by Joule's effect at the surface of the metal is on average
\be
P_{\rm joule}= \left \langle j_{\rm ind}(0,t) e(0,t)\right \rangle= \frac{J_0^2}{2m^3}.
\ee
Let us now assume that dissipation acts adiabatically over times $t_{\rm dis}\gg 1/m$, then the average energy decreases as
\be
\frac{d\langle E\rangle}{dt}= \frac{1}{2m^3\sigma}\frac{dJ_0^2}{dt},
\ee
i.e.  the amplitude of the current inside matter varies  in time due to dissipation.

As we consider the balance equation at $z=0$ we have
\be
\langle \vec \nabla \vec P_{\rm Poynting}\rangle = \left \langle
\frac{d e}{dz}(0,t) b(0,t)+ \frac{db}{dz}(0,t)e(0,t)\right \rangle,
\ee
where the Bianchi identity gives us that
\be
b_y= -\frac{J_0}{m^2}\cos m(t-z) + \frac{J_0}{m\sqrt{m\sigma}}\sin\left ( m(t-z) -\frac{\pi}{4}\right ).
\ee
Notice that the non-propagating term of $e(z,t)$ in $\cos mt$ corresponds to a term with a Fourier transform proportional to $\delta (k)$ as a result we can also write $b_y= \frac{k}{\omega} e(k,\omega)$ where the Fourier transform of the non-propagating part is
\be
e(\omega,k) \supset \frac{J_0}{2m^2}\left (\slashed{\delta}(\omega-m) + \slashed{\delta} (\omega + m)\right )\slashed{\delta}(k),
\ee
whose contribution does not appear in $b_y(k,\omega)$ as $k\slashed{\delta}(k)\equiv 0$. Hence only the propagating part of $e$ contributes to $b_y$ as $k=\pm m$.
We then find that $\langle b(0,t) \partial_z e(0,t)\rangle =0$ and $\langle \vec \nabla \vec P_{\rm Poynting}\rangle= \langle \partial_z b(0,t) e(0,t)\rangle $ which gives
\be
\langle \vec \nabla \vec P_{\rm Poynting}\rangle =-\frac{J_0^2}{2\sqrt 2 m^2 \sqrt{m\sigma}}.
\ee
Similarly we find that
the axion injects energy in the system as
\be
\langle j_{\rm axion}(0,t) e(0,t)\rangle= \frac{J_0^2}{2\sqrt 2 m^2 \sqrt{m\sigma}},
\ee
implying that the radiated energy balances the axionic injection
\be
\langle \vec \nabla \vec P_{\rm Poynting}\rangle + \langle j_{\rm axion}(0,t) e(0,t)\rangle =0.
\ee
This implies that the change of amplitude of the current satisfies
\be
\frac{d\langle E\rangle}{dt}= -\langle j_{\rm ind}(0,t) e(0,t)\rangle,
\ee
where the variation of the electromagnetic energy at the surface of the plaque is due to the Joule effect from the induced current. This would give
\be
J_0^2(t)= J_0^2(0) e^{-\sigma t},
\ee
corresponding to a dissipation time $t_{\rm dis}= 1/\sigma$. Typically for Copper, we have $\sigma=10^{-8} \;{\rm GeV}$ implying that the dissipation time appears to be much shorter than the oscillation time $1/m$ as long as $m\le 10$ eV. In this case, the previous calculation does not apply and one must take into account both the time evolution of the axions and the photons.

\subsection{Axion electrodynamics in the plaque}

Let us now go beyond the simple hypothesis that $\phi_0$ is constant in the plaque. The Joule's effect implies that this cannot be the case. We must analyse the coupled Klein-Gordon equation following the line of~\cite{Ahonen:1995ky}
\be
\ddot \phi -\Delta \phi +m^2 \phi= -\frac{B}{M}\dot a
\ee
where $e=-\dot a$. The equation for the electric field becomes
\be
\Delta a - \ddot a= -\frac{B}{M} \dot \phi + \sigma \dot a,
\ee
where the term $j_\sigma= \sigma e$ is the conduction current.

We can analyse the modes of this system in the plaque by going to Fourier space. Having two equations with no sources, the non-trivial modes are obtained by requiring that the determinant of the system vanishes. This gives
\be
(\omega^2 -k^2 +i\sigma \omega) (\omega^2 - k^2 -m^2)= \frac{B^2 \omega^2}{M^2}.
\ee
As we are interested in $B/M \ll m$, the mixing is very small implying that there are two branches of solutions. The first branch corresponds to
\be
\omega^2 -k^2 +i\sigma \omega\approx 0,
\ee
corresponding to two modes
\be
\omega_1\approx -i\sigma, \ \omega_2\approx -i \frac{k^2}{\sigma},
\ee
where we have assumed that $k\ll \sigma$. Both modes are decaying modes in time with no time oscillations. So they do not correspond to the case of an initial oscillatory axion being affected by Joule's effect. The mode $\omega_1$ decays very fast whereas the second one has a long lifetime.

The more interesting modes which would correspond to an initially oscillating axions are on the second branch close to $\omega^2=k^2 +m^2$. The modes are then
\be
\omega_\pm= \pm \left ( (k^2+m^2)^{1/2}+ \frac{m^2}{2 \sigma^2}\frac{B^2}{M^2}\frac{1}{\sqrt{k^2+m^2}} -\frac{B^2}{2M^2}\frac{i}{\sigma}\right )
\ee
The physical mode is the decreasing one $\omega_+$ whose imaginary part is in $-\frac{B^2}{2M^2}\frac{i}{\sigma}$.
Notice that when $\sigma$ is very large, this gives a very large lifetime. This is the surprising result of this analysis. The mode $\omega_1$ corresponds to the intuition that the Joule's effect withdraws the energy from a finite source and therefore depletes the axion and the electric field very rapidly. The mode $\omega_+$ corresponds to the fact that the Joule's effect removes energy from the system but the axion keeps replenishing. The two effects compete but the second one with a long lifetime is the dominant effect.

As a result, the analysis in the plaque confirms that our initial hypothesis of a constant amplitude for the axion is justified as the variation time due to the Joule effect is very large. So we are entitled to trust our analysis and keep that $J_0$ is constant on the time scale of the experiment.

\section{Comparison with the Lifschitz theory}

For a metal the Casimir effect depends on the conductivity too.  In this case, the pressure separates in the electric pressure
\be
P_z^+=  -\int \slashed{d}\omega \slashed{d}^2 p_\parallel  \frac{\Delta}{{\left (\frac{1+\frac{\xi}{\Delta}}{1-\frac{\xi}{\Delta}}\right )^2 e^{2\Delta d}-1}},
\ee
coming from the $TE$ modes with two plates with the permittivity $\epsilon$. We have introduced $\Delta= (p^2_\parallel +\omega^2)^{1/2}$. This  integral can be written as
\be
P_z^+=  -\int \slashed{d}\omega \slashed{d}^2 p_\parallel  \frac{\Delta}{{r_{TE}^{-2} e^{2\Delta d}-1}},
\ee
where $r_{TE}$ is the reflection coefficient for the $TE$ modes evaluated for imaginary frequencies $r_{TE}(i\omega)= \frac{1-\frac{\xi}{\Delta}}{1+\frac{\xi}{\Delta}}$ where $\xi=(p_\parallel^2 +\omega^2 \epsilon(i\omega))^{1/2}$ depends on the permittivity for imaginary frequencies
\be
\epsilon(i\omega)= 1 + \frac{\omega^2_{\rm Pl}}{\omega(\omega+\gamma)}.
\ee
The magnetic pressure is given by
\be
P_z^-=  -\int \slashed{d}\omega \slashed{d}^2 p_\parallel  \frac{\Delta}{{\left (\frac{1+\frac{\xi}{\epsilon(i\omega)\Delta}}{1-\frac{\xi}{\epsilon(i\omega)\Delta}}\right )^2 e^{2\Delta d}-1}},
\ee
coming from the $TM$ modes with two plates with the same permittivity $\epsilon$. Notice that the same integral can be written as
\be
P_z^-=  -\int \slashed{d}\omega \slashed{d}^2 p_\parallel  \frac{\Delta}{{r_{TM}^{-2} e^{2\Delta d}-1}},
\ee
where $r_{TM}$ is the reflection coefficient for the $TM$ modes evaluated for imaginary frequencies.
In the ideal case where $r_{TE}=r_{TM}=1$, the electric and magnetic contributions to the Casimir pressure are equal and the total pressure  reduces to
\be
P_z= -2\int \slashed{d}\omega \slashed{d}^2 p_\parallel  \frac{\Delta}{e^{2\Delta d}-1},
\ee
which can be evaluated as
\begin{widetext}
\be
P_z= -\frac{2\times 4\pi}{16(2\pi)^3 d^4}\int_0^\infty dx \frac{x^3}{e^x-1}= - \frac{2\times 4\pi}{16(2\pi)^3 d^4}\Gamma(4) \zeta^(4)= -\frac{\pi^2}{240 d^4}.
\ee
\end{widetext}
This is the usual Casimir pressure in the ideal case.
At finite temperature $T$, the Lifschitz formulae must be computed by replacing the integral $\int \slashed{d} \omega \to T \sum_{n>0} $ over the Matsubara frequencies where
$\omega_n= 2\pi nT$.
As the reflection coefficients are less than unity, the effect of the conductivity of the real plates is to reduce the quantum pressure between the plates. {Finally, let us notice that when $\omega$ becomes larger than $\omega_{\rm Pl}$, the reflection coefficients converge to zero implying that the pulsation  integral giving rise to the Casimir pressure has a natural cut-off scale of order $\omega_{\rm Pl}$ above which the integral receive no contribution. Similarly when the transverse momenta $p_\parallel$ are much larger than $\omega$, the TE reflection coefficient vanishes too. In all cases, when the quantised pulsation
$\omega_n \sim \sqrt{ p^2_\parallel+\frac{n^2\pi^2}{d^2}}$ exceeds the plasma frequency, in particular for large $p_\parallel$, the plaques become transparent implying that $\epsilon=1$ and the absence of Casimir pressure. As a result the plasma frequency serves as a UV cut-off for energies and momenta for the Casimir effect in the presence of metallic plates.  }

\section{Quantum Effects}
\label{app:quantum}
The contribution of the axion to the quantum pressure can be evaluated by considering the corrections of the photon propagator due to the axion.
For each photon polarisation, let us consider the vacuum energy per unit surface  which behaves like
\be
\f{E_{\rm vac}}{S}= \int \slashed{d} \omega \slashed{d}^2 p_\parallel \omega^2 \Delta(\omega, p_\parallel)
\ee
where the photon propagator in the cavity with resonances at $\omega= \omega_n(p_\parallel) \sim p^2_\parallel +\frac{n^2\pi^2}{d^2}$, where $\vec p_\parallel$ is the transverse momentum to the plates, is given by
\be
\Delta(\omega, p_\parallel)\sim \sum_n \frac{i}{\omega^2 - \omega^2_n(p_\parallel) +i\epsilon}.
\ee
A contour integral in the lower half plane gives the usual $\frac{1}{2}\sum_n \int \slashed{d}^2 p_\parallel \omega_n(p_\parallel)$ before the axionic corrections. As the photon energy momentum tensor is not corrected by the axion coupling due to its topological nature, we just have to correct the propagator and average over the fast axion oscillations. The first correction to the vacuum energy appears after two axionic insertions and gives a term in
\begin{widetext}
\be
\f{\delta E_{\rm vac}}{S}\sim  m^2 \frac{\phi_0^2}{2M^2} \int \slashed{d} \omega \slashed{d}^2 p_\parallel  p^2_\parallel \Delta(\omega, p_\parallel) ( \Delta(\omega+m, p_\parallel)+ \Delta(\omega-m, p_\parallel)) \Delta(\omega, p_\parallel)
\ee
\end{widetext}
where each vertex bring a spatial and a time derivative. The time derivatives lead to the prefactor in $m^2$ coming from the derivative of the axion field and the spatial derivative is in $p_\parallel$. In this expression, we are only interesting in the order of magnitude. A more precise calculation using field theoretic methods from first principle is in progress. Notice that the axionic insertions shift the energy in the propagator $\omega \to \omega \pm m$. The integral is dominated by the resonances at $\omega_n$ and gives a leading contribution
\begin{widetext}
\be
\f{\delta E_{\rm vac}}{S}\sim i m^2 \frac{\phi_0^2}{4M^2}\sum_n  \int \slashed{d} \omega \slashed{d}^2 p_\parallel \omega^2  p^2_\parallel \frac{1}{(\omega^2 - \omega^2_n(p_\parallel) +i\epsilon)^2} \frac{1}{\omega_n^2(p_\parallel)-\frac{m^2}{4}}
\ee
\end{widetext}
The same contour integral in the lower half plane gives a contribution in
\be
\f{\delta E_{\rm vac}}{S}|\sim  m^2 \frac{\phi_0^2}{4M^2}\sum_n  \int \slashed{d} \omega \slashed{d}^2 p_\parallel  p^2_\parallel \frac{1}{\omega_n(p_\parallel)} \frac{1}{\omega_n^2(p_\parallel)-\frac{m^2}{4}}
\ee
which is an integral which can be calculated in dimensional regularisation
\begin{widetext}
\be
\f{\delta E_{\rm vac}}{S}\sim  m^2 \frac{\phi_0^2}{4M^2}\sum_n  \int  \slashed{d}^2 p_\parallel  p^2_\parallel \frac{1}{(p^2_\parallel +\frac{n^2\pi^2}{d^2})^{3/2}}\sim
 m^2 \frac{\phi_0^2}{16\pi M^2} \frac{\Gamma(-\frac{3}{2})}{\Gamma(\frac{1}{2}) } \sum_n \frac{n\pi}{d}
\ee
\end{widetext}
when $md\ll 1$. This gives
\be
\f{\delta E_{\rm vac}}{S} \sim  m^2 d^2  \frac{\phi_0^2}{16 M^2} \frac{\Gamma(-\frac{3}{2})}{\Gamma(\frac{1}{2})}\zeta(-1) \frac{1}{d^3}
\ee
and dimensionally we have
\be
\frac{\delta E_{\rm vac}}{E_{\rm vac}} \sim  m^2 d^2  \frac{\phi_0^2}{ M^2}
\ee
as $\f{E_{\rm vac}}{S}\sim 1/d^3$.

\end{document}